\documentclass[%
 reprint,
]{revtex4-2}

\usepackage{graphicx}
\usepackage{dcolumn}
\usepackage{bm}
\usepackage{xcolor}
\usepackage{booktabs}
\usepackage{array}
\usepackage{multirow}
\usepackage{algorithm}
\usepackage{algpseudocode}
\usepackage{comment}
\usepackage{threeparttable}
\usepackage{adjustbox}
\usepackage{subcaption}

\newcommand{\metric}[1]{\makebox[1.65cm][c]{#1}}

\definecolor{softviolet}{RGB}{150,150,255}
\newcommand{\fasercal}{\textsc{FASERCal}}
\newcommand{\threedcal}{\textsc{3DCal}}
\newcommand{\ahcal}{\textsc{AHCAL}}
\newcommand{\ecal}{\textsc{ECAL}}

\newcommand{\scratchmodel}{Scratch}
\newcommand{\maemodel}{MAE}
\newcommand{\maerelmodel}{MAE+Rel}

\begin{document}

\title{Towards foundation-style models for energy-frontier heterogeneous\\neutrino detectors via self-supervised pre-training}

\author{Saúl Alonso-Monsalve}\email{salonso@ethz.ch}
\author{Fabio Cufino}\email{fcufino@ethz.ch}
\author{Umut Kose}
\author{Anna Mascellani}
\author{André Rubbia}
\affiliation{%
 IPA, ETH Zürich, Otto Stern Weg 5, Zurich, 8093, Switzerland.
}

\date{\today}

\begin{abstract}
Accelerator-based neutrino physics is entering an energy-frontier regime in which interactions reach the TeV scale and produce exceptionally dense, overlapping detector signatures. Event interpretation is impractical for conventional reconstruction and challenging for supervised models trained from scratch, particularly when labelled data are scarce and analyses span diverse objectives. We present a sparse Vision Transformer framework for learning reusable representations from heterogeneous detector data. Self-supervised pre-training combines masked-autoencoder reconstruction with relational voxel-level objectives for hierarchy, ghost and particle identification, followed by joint fine-tuning across classification and regression tasks. On simulated \fasercal{} events, pre-training improves neutrino-flavour and charm-quark identification, momentum regression and vertex reconstruction, with relational objectives providing further gains in topologically complex channels. Attribution and representation diagnostics show a more structured latent space, while detector-subsystem ablations recover physically plausible channel-dependent roles. With roughly $10^3$ labelled events, the pre-trained encoder matches the flavour-classification performance of scratch training with an order of magnitude more data. Cross-domain transfer is observed on public plastic-scintillator and liquid-argon benchmarks. A matched alternative-generator stress test shows stable flavour and kinematic performance while exposing charm tagging as generator-sensitive. These results support self-supervised multimodal pre-training as a route towards reusable detector representations. We use ``foundation-style'' in this restricted sense, rather than claiming a completed general-purpose detector foundation model.
\end{abstract}


\maketitle

\section{\label{sec:intro}Introduction}

Accelerator-based neutrino physics is entering the energy frontier. Forward collider-neutrino programmes extend measurements into the TeV regime, enabling neutrino-interaction studies and cross-section measurements at previously unexplored energies while broadening the case for collider neutrinos as a high-energy physics programme~\cite{abreu2020detecting,abreu2023firstdirect,abraham2024crosssections,cruzmartinez2024lhcneutrinoion, FASER:2024ref, SNDLHC2023}. Yet the same regime produces detector data of exceptional complexity. Interaction topologies become highly collimated, particle multiplicities rise, electromagnetic and hadronic activity overlap strongly, and information must be integrated across heterogeneous detector systems. Machine learning is already widely used across particle physics for detector inference, anomaly detection, or event reconstruction, and that broader landscape has been reviewed extensively~\cite{radovic2018mlfrontiers,coelho2021quantization,karagiorgi2022mlnewphysics,govorkova2022autoencoders,belis2024quantum,pata2024particleflow}. In energy-frontier neutrino detection, however, the challenge is not simply whether learned models outperform existing pipelines, but whether any practical analysis of these events is feasible without them.

\begin{figure*}[htpb]
\centering

\begin{subfigure}[t]{0.49\textwidth}
\centering
\text{ Event A - CC $\nu_\mu$, $E_\nu = 353$ GeV}\\[+2pt]
\includegraphics[width=\linewidth]{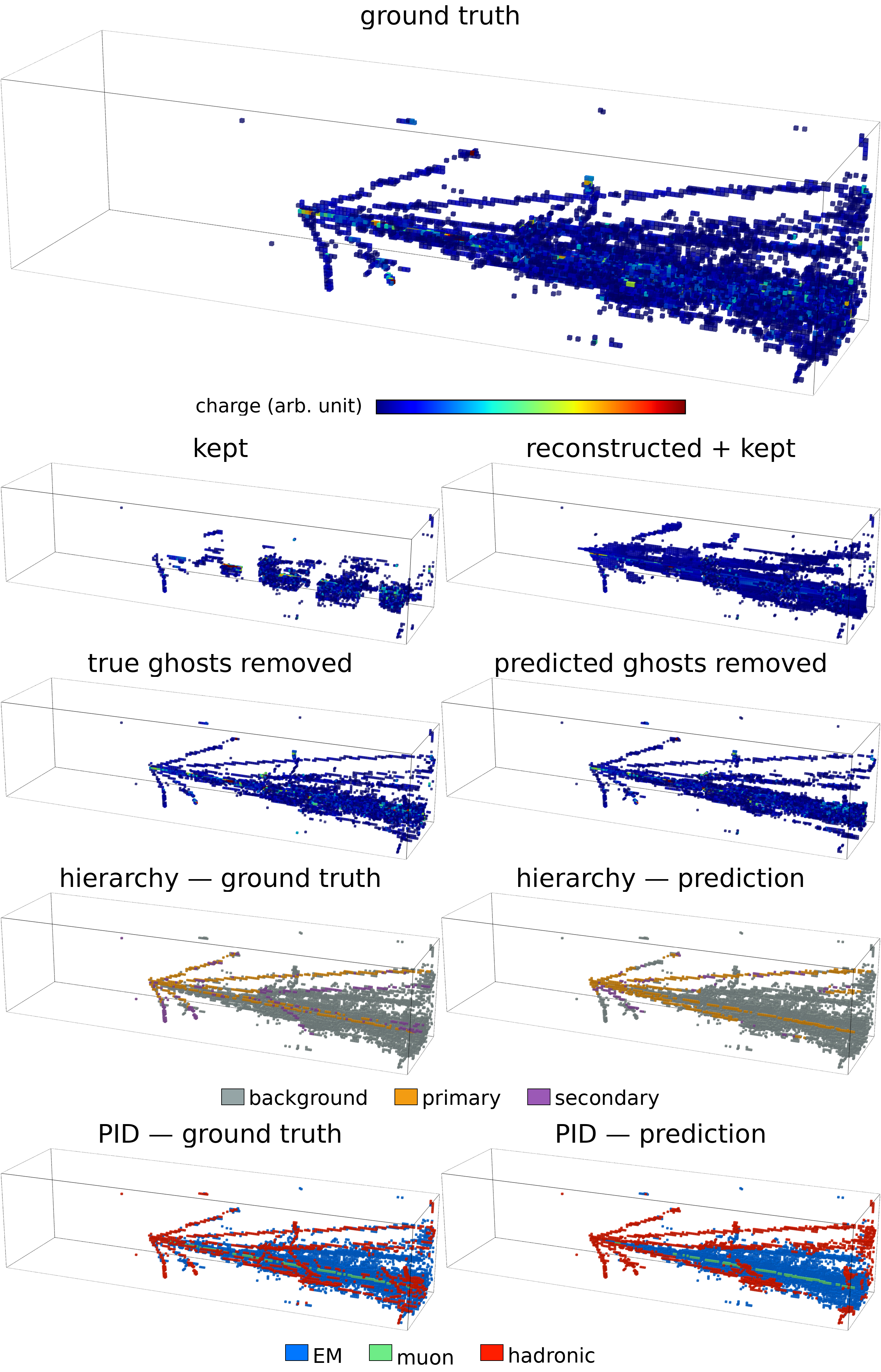}
\end{subfigure}
\hfill
\begin{subfigure}[t]{0.49\textwidth}
\centering
\text{ Event B - CC $\nu_\mu$, $E_\nu = 1380$ GeV}\\[+2pt]
\includegraphics[width=\linewidth]{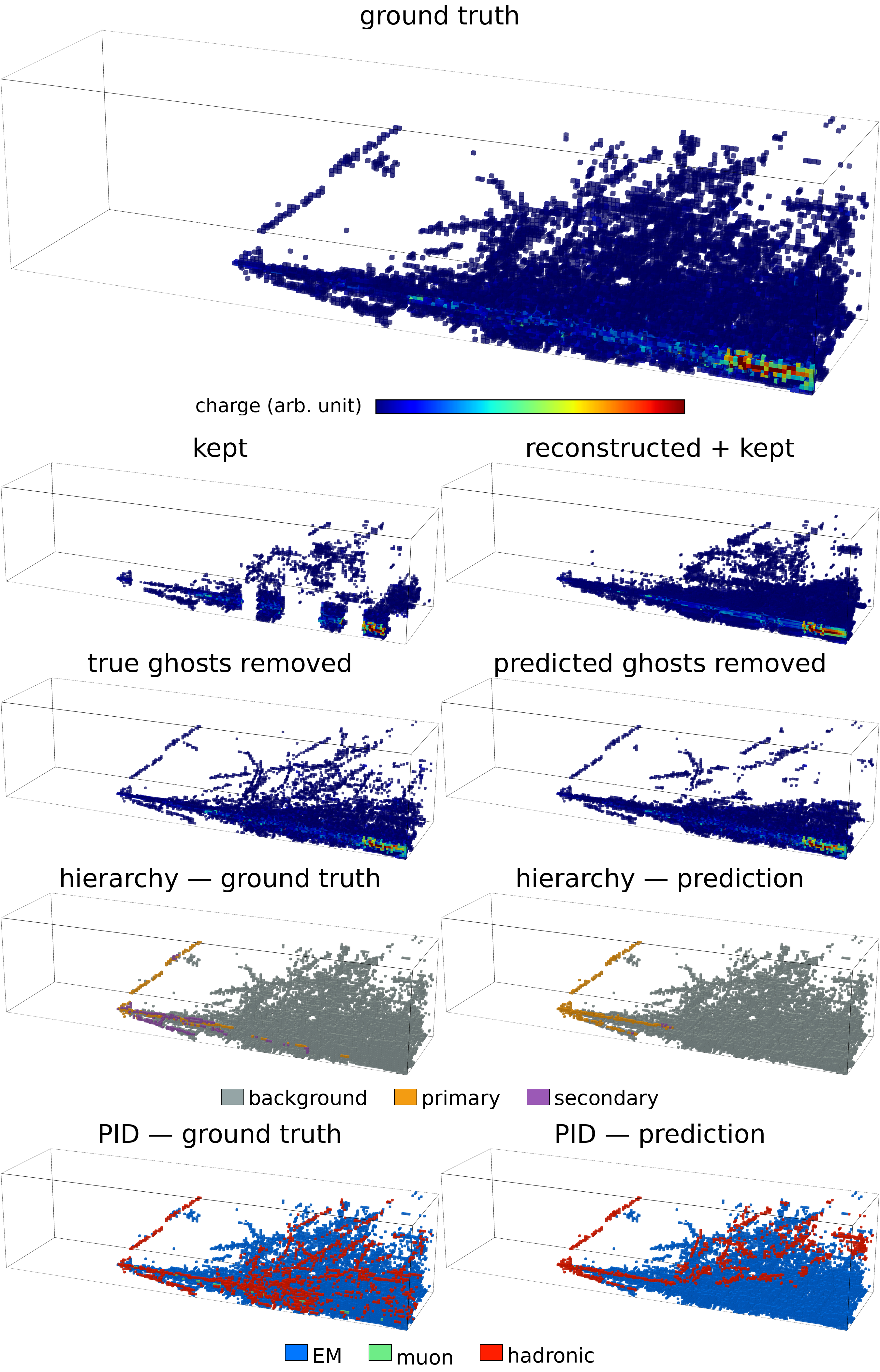}
\end{subfigure}

\vspace{4pt}

\caption{\textbf{Example events from the test set illustrating masked reconstruction and relational tasks in the \threedcal{} detector.}
For each event, the top panel displays the ground-truth detector readout.
The second row presents the visible voxels in the masked input (25\% \textit{kept}) and the reconstructed output combined with the visible voxels (\textit{reconstructed + kept}).
The third row compares ghost-hit removal using ground truth and model predictions.
The fourth row illustrates interaction hierarchy labelling (background, primary, secondary).
The bottom row presents particle identification (electromagnetic, muon, hadronic).
For the hierarchy and particle-identification panels, each voxel is assigned to its dominant contributing class for visualisation.
}

\label{fig:event_examples}

\end{figure*}

Within accelerator neutrino experiments, convolutional, graph-based and transformer-like models have been shown to improve event classification, semantic segmentation, or clustering on simulated detector data~\cite{aurisano2016cvn,acciarri2017microboonecNN,psihas2020review,abi2020dunecvn,domine2020sparse,abratenko2021sparsessnet,PhysRevD.103.032005,drielsma2020grappa,alonsoMonsalve2024overlap}. Most of the literature, however, concerns lower-energy environments, individual detector subsystems, or task-specific supervised models trained from scratch, often in settings where conventional reconstruction remains plausible. The regime considered here is qualitatively different: event topologies are so dense, collimated and overlapping that the challenge is not to refine an existing analysis chain, but to construct a viable one.

The \fasercal{} concept, a proposed upgrade for the FASER experiment at CERN, provides a stringent case study for this problem~\cite{CERN-FASER-NOTE-2026-004}. Its highly granular \threedcal{} comprises over 460,000 readout voxels, only a fraction of which are active in a given event, and is followed by electromagnetic and hadronic calorimeters and a muon spectrometer. The forward neutrino beam contains $\nu_e$, $\nu_\mu$, and $\nu_\tau$ components, and the analyses considered here target both charged-current (CC) and neutral-current (NC) interactions. A model must therefore integrate sparse three-dimensional volumetric inputs with heterogeneous auxiliary streams of different dimensionality. The resulting events, illustrated in Fig.~\ref{fig:event_examples}, contain dense shower cores, extended secondary tracks, partial containment and deeply ambiguous local configurations. In this setting, machine-learning-based analysis is not an optional enhancement to an otherwise tractable reconstruction problem; it is a mandatory prerequisite for practical event-level physics extraction.

This is precisely the setting in which representation learning becomes compelling. High-fidelity simulation provides abundant detector responses, but labelled targets remain task-specific and expensive to produce. Masked modelling has emerged as a strong route to transferable representations in language, vision, and structured data~\cite{devlin-etal-2019-bert,he2022masked,Xie_2022_CVPR,pmlr-v162-baevski22a,Yu_2022_CVPR,hou2022graphmae}. In neutrino and particle physics, recent studies have begun exploring transfer learning, domain adaptation, contrastive learning, and masked point modelling~\cite{Chappell2022Application,Babicz2022Adversarial,Sagar2025Adapting,Albert2025Deep,Wilkinson2025Contrastive,Yu:2025t0,Young2026Particle,Bonilla2026Transfer}. What remains missing is a unified approach for learning reusable models for genuinely heterogeneous detectors in the extreme topological regime of energy-frontier neutrino interactions.

Our aim is to take a concrete step towards foundation-style models for neutrino detector data. We do not claim a completed general-purpose model; instead, we target the core ingredients such a model requires: broad self-supervised pre-training, reused across several downstream tasks, strong performance when only a few hundred to a few thousand labelled events are available, and transfer across detector technologies and energy scales. In this manuscript, ``foundation-style'' therefore has an operational meaning: a single reusable encoder is trained with objectives independent of downstream task labels on a large detector sample, adapted to several supervised physics tasks, tested under limited-label conditions, probed with a matched alternative-generator sample, and reused outside the source detector. A full detector foundation model would require broader detector coverage, stronger systematic validation, and ultimately experimental-data calibration. We address this with a sparse vision-transformer (ViT)-like~\cite{dosovitskiy2021vit} framework that combines masked-autoencoder (MAE)~\cite{he2022masked} pre-training with relational voxel-level objectives available from simulation. The encoder is later fine-tuned on flavour identification, charmed-quark identification, event kinematics and vertex reconstruction, stress-tested on matched GENIE and NuWro samples, and evaluated for transfer on public datasets outside the source domain.

This paper makes three main contributions.
\begin{enumerate}
    \item We introduce a sparse encoder for heterogeneous detector data that combines sparse convolutional patch embeddings, module-aware self-attention, and Perceiver-IO fusion across calorimetric and tracking streams.
    \item We formulate a multimodal pre-training strategy that augments masked reconstruction with relational voxel-level targets (ghost identification, interaction hierarchy and particle category), and show in the in-domain comparisons that include all three baselines that this composite objective improves downstream performance beyond MAE-only pre-training, with the largest gains in the most challenging channels.
    \item We demonstrate that the learned representation improves performance and data efficiency across a multi-task fine-tuning suite, and transfers beyond the source domain to publicly available benchmarks spanning different detector technologies and energy regimes. We also include a matched GENIE-to-NuWro evaluation of the final \maerelmodel{} to probe generator dependence without retraining, and report the cross-domain transfer summaries for all three initialisation strategies.
\end{enumerate}

We evaluate this framework first on simulated events from the \fasercal{} concept as an energy-frontier case study, including a matched GENIE-to-NuWro stress test, and then on public transfer benchmarks covering a variety of detector technologies and energy scales. The Results section therefore moves from pre-training behaviour to downstream classification, regression, attribution and representation diagnostics, data efficiency, and a combined robustness-and-transfer evaluation. The Discussion section interprets what these findings imply for representation learning in detector physics, and the Methods describe the case-study data, architecture, training strategy, and transfer setup.

\section{\label{sec:results}Results}

All downstream comparisons use the same architecture under three initialisation strategies:
\begin{enumerate}
    \item \textbf{\scratchmodel}: the fine-tuning architecture with standard random initialisation of all trainable parameters.
    \item \textbf{\maemodel}: the same downstream architecture and randomly initialised task heads, with the encoder initialised from masked-reconstruction pre-training.
    \item \textbf{\maerelmodel}: the same downstream architecture and randomly initialised task heads, with the encoder initialised from the full pre-training objective that combines masked reconstruction with voxel-level relational tasks.
\end{enumerate}

Throughout the downstream results, \scratchmodel{}, \maemodel{} and \maerelmodel{} denote the fully fine-tuned multi-task models, not frozen encoders.

\subsection{\label{sec:pretraining_results}Pre-training}

The pre-training tasks are intended to shape the latent representation rather than to serve as end products, so we present them primarily as qualitative evidence. Figure~\ref{fig:event_examples} shows masked-reconstruction examples on test events. Even with 75\% of occupied patches masked, the model recovers broad shower envelopes, elongated track-like structures and coherent energy flow into missing regions. The reconstructions are not exact, particularly in dense shower cores and around fine secondary structures, and we do not interpret them as precise generative predictions. Their value is diagnostic: the encoder is forced to infer non-local spatial correlations rather than copying visible voxels. The relational objectives are also shown in Fig.~\ref{fig:event_examples}. The model predicts ghost labels, hierarchy labels and particle-category labels on kept patches, again on test events. Here, ghost voxels are reconstructed deposits with no matched true particle, hierarchy labels distinguish background, primary and secondary activity, and particle-category labels separate electromagnetic, muonic and hadronic deposits.
These targets are especially demanding: a single reconstructed voxel can receive contributions from multiple true particles, yielding soft class mixtures rather than hard one-hot assignments. Agreement with ground truth is strongest in well-separated tracks and along the dominant shower axes, while the densest overlapping regions remain the most difficult. This is consistent with the detector challenge described in Ref.~\cite{CERN-FASER-NOTE-2026-004}: dense occupancy and view ambiguities create ghost activity and complicate local pattern assignment. Taken together, the reconstruction and relational results in Fig.~\ref{fig:event_examples} suggest that the encoder captures meaningful spatial and semantic correlations before any downstream fine-tuning is applied.

\subsection{\label{sec:classification_results}Classification and regression}

\begin{figure*}[htpb]
  \centering
  \includegraphics[width=0.98\linewidth]{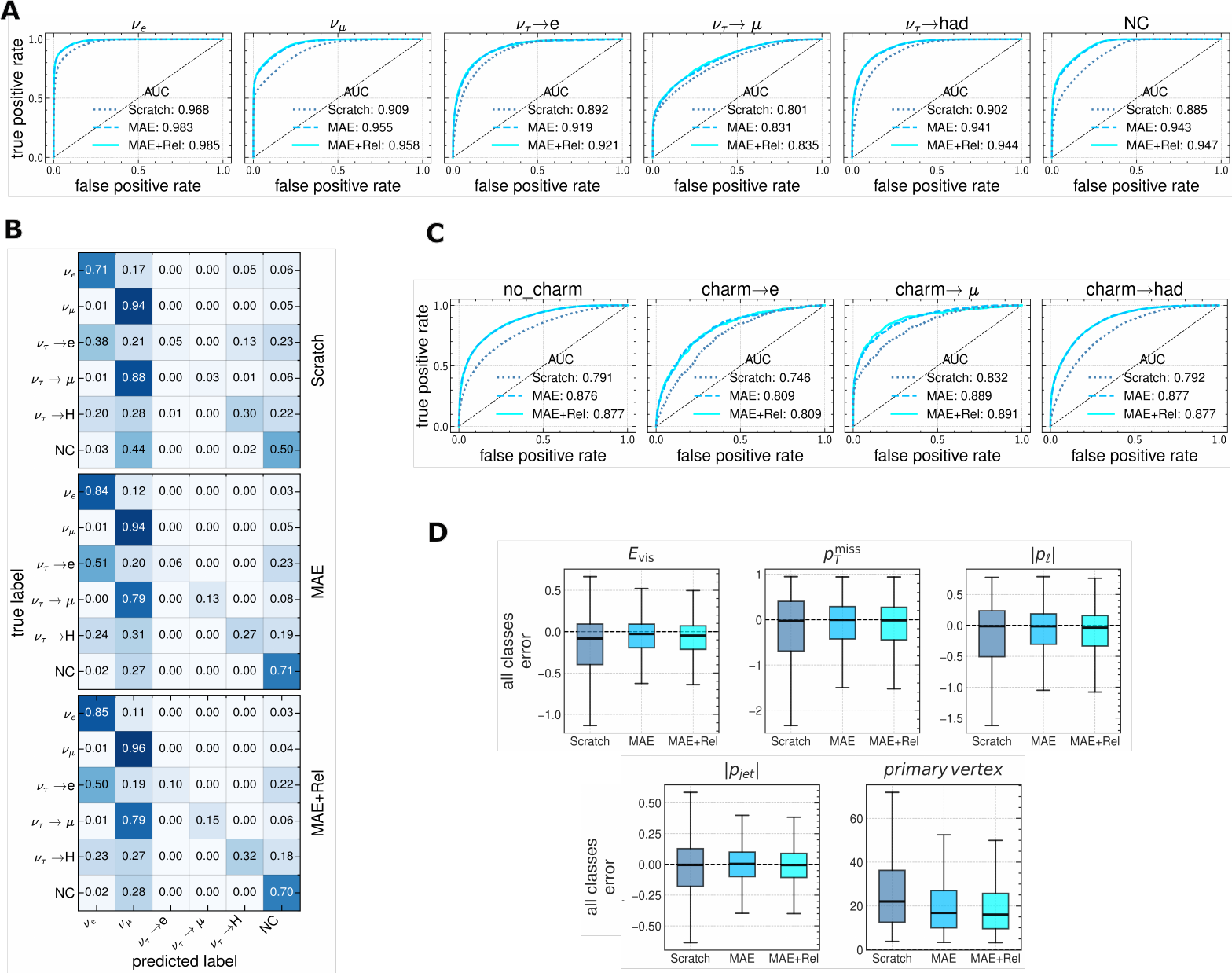}
  \caption{\textbf{Classification and regression performance.}
  \textbf{A} One-vs-rest ROC curves for the six flavour categories, comparing \scratchmodel{}, \maemodel{} and \maerelmodel{}.
  \textbf{B} Row-normalised flavour confusion matrices using per-class thresholds that maximise the figure of merit; events passing multiple thresholds are assigned to the highest score.
  \textbf{C} One-vs-rest ROC curves for charm categories.
  \textbf{D} All-class regression-error summary for the selected event sample, shown for $E_{\mathrm{vis}}$, $p_T^{\mathrm{miss}}$, $|p_\ell|$, $|p_{\mathrm{jet}}|$ and the primary-vertex error. For the momentum and energy observables, errors are fractional residuals $(x^{\mathrm{true}}-x^{\mathrm{reco}})/x^{\mathrm{true}}$; for the primary vertex, the error is $d_{\mathrm{PV}}=\|\vec{x}_{\mathrm{PV}}^{\mathrm{true}}-\vec{x}_{\mathrm{PV}}^{\mathrm{reco}}\|$ in mm. Boxplots show the median, interquartile range and whiskers extending to approximately 95\% of the selected sample. The corresponding purity, efficiency and figure-of-merit scans, together with class-resolved regression boxplots, are shown in Supplementary Information~\ref{app:detailed_clsreg}.
  Definitions: ROC: receiver operating characteristic; AUC: area under the curve; NC: neutral current; $\mathrm{had}$: hadronic; FOM: figure of merit ($S/\sqrt{S+B}$); $S$ and $B$ denote expected signal and background yields.}
  \label{fig:classification_regression_results}
  \label{fig:classification_results}
\end{figure*}

The downstream classification and regression results are summarised together in Fig.~\ref{fig:classification_regression_results}. The figure combines the most diagnostic classification panels with an all-class regression summary, while the operating-point scans and class-resolved regression boxplots are provided in Supplementary Information~\ref{app:detailed_clsreg}. Performance is evaluated with one-vs-rest ROC curves, row-normalised confusion matrices, and the figure of merit $\mathrm{FOM}=S/\sqrt{S+B}$, where $S$ and $B$ denote the expected signal and background yields after thresholding. For flavour classification, CC $\nu_\tau$ events are split into $\tau\!\to\! e$, $\tau\!\to\! \mu$ and $\tau\!\to\! \mathrm{had}$ according to whether an electron or muon appears among the primary tau-decay products; all remaining CC $\nu_\tau$ events are assigned to the hadronic class. For charm classification, events are assigned to charm$\to\mu$ if any primary charm-decay product is a muon, to charm$\to e$ otherwise if any primary charm-decay product is an electron, and to charm$\to\mathrm{had}$ for the remaining charm decays. The overall pattern is clear. \maemodel{} already improves the dominant channels, while \maerelmodel{} provides the largest gains for the less abundant and more topologically complex signatures.

For the dominant channels, the gains are consistent. The $\nu_e$ CC area under the receiver-operating-characteristic curve increases from 0.968 for \scratchmodel{} to 0.983 for \maemodel{} and 0.985 for \maerelmodel{}, the $\nu_\mu$ CC area under the curve increases from 0.909 to 0.955 and 0.958, and the neutral-current area under the curve increases from 0.885 to 0.943 and 0.947. The row-normalised confusion matrices also become more diagonal for the common categories. For example, the $\nu_e$ CC diagonal entry rises from 0.71 to 0.84 and 0.85, while the neutral-current diagonal rises from 0.50 to 0.71 and 0.70. This suggests that masked reconstruction effectively captures the global event morphology needed for the dominant classes.

The most consequential gains appear in the tau-neutrino channels. The threshold scans supporting the quoted maximum-FOM values are shown in Supplementary Figure~\ref{fig:classification_supplement}. For $\nu_\tau$ CC $\rightarrow \mathrm{had}$, the area under the curve rises from 0.902 to 0.941 and 0.944, and the maximum figure of merit increases from 1.58 to 4.43 and 4.58. For $\nu_\tau$ CC $\rightarrow e$, the area under the curve rises from 0.892 to 0.919 and 0.921, while the maximum figure of merit increases from 0.38 to 1.03 and 1.35. The muonic tau channel shows a similar trend, with area under the curve improving from 0.801 to 0.831 and 0.835 and the maximum figure of merit from 1.25 to 1.79 and 1.82. The confusion matrices show that these channels remain the most difficult, with substantial leakage into $\nu_e$ CC, $\nu_\mu$ CC, and neutral-current categories, but the pre-trained models separate them more cleanly than \scratchmodel{}, especially once the relational objective is included. These are precisely the channels in which dense overlap, secondary activity and partial containment complicate classification, so the quantitative pattern is important: the benefit of pre-training is most pronounced where event interpretation is hardest.

Charmed-quark classification shows the same structure. The no-charm category is already separated well by all models, but the less abundant charm decays benefit substantially from pre-training. For charm $\rightarrow \mu$, the area under the curve increases from 0.832 to 0.889 and 0.891, and the maximum figure of merit rises from 6.97 to 7.61 and 7.90. For charm $\rightarrow \mathrm{had}$, the area under the curve increases from 0.792 to 0.877 for both pre-trained variants, while the maximum figure of merit rises from 27.10 to 37.27 and 37.74. Even charm $\rightarrow e$, which remains challenging, shows an area-under-the-curve increase from 0.746 to 0.809 and a maximum-figure-of-merit increase from 1.75 to 2.08 and 2.20. Overall, the relational objectives do not simply nudge the model upwards. They disproportionately improve the lower-yield channels that are most critical for the detector's physics reach.

The same trend extends beyond classification. Panel~D of Fig.~\ref{fig:classification_regression_results} summarises the regression errors for the event visible energy $E_{\mathrm{vis}}$, the missing transverse momentum $p_T^{\mathrm{miss}}$, the magnitudes of the primary-lepton and hadronic-jet momenta $|p_\ell|$ and $|p_{\mathrm{jet}}|$, and the primary-vertex position after pooling the selected events across flavour categories. Primary-vertex performance is summarised with the 3D error $d_{\mathrm{PV}}=\|\vec{x}_{\mathrm{PV}}^{\mathrm{true}}-\vec{x}_{\mathrm{PV}}^{\mathrm{reco}}\|$. Performance is best assessed through shifts in the medians and reductions in spread across the boxplots, with class-resolved regression boxplots shown in Supplementary Figure~\ref{fig:regression_supplement}.

The clearest and most uniform effect appears in primary-vertex reconstruction. In the all-class summary, the $d_{\mathrm{PV}}$ distributions for the pre-trained models are shifted towards smaller errors than \scratchmodel{} and their interquartile ranges are reduced. The class-resolved boxplots in Supplementary Figure~\ref{fig:regression_supplement} show that this behaviour is not driven by a single easy topology, but appears across the common $\nu_e$ CC and $\nu_\mu$ CC channels as well as the more difficult tau and neutral-current samples. Between the two pre-trained variants, \maerelmodel{} is typically the most compact or among the most compact distributions.

The kinematic targets show a more heterogeneous but still favourable pattern. For $E_{\mathrm{vis}}$ and $|p_{\mathrm{jet}}|$, the pre-trained models generally move the medians closer to zero and reduce the spread, with the clearest gains in the charged-current channels and in $\nu_\tau$ CC $\rightarrow \mathrm{had}$, where jet reconstruction is especially challenging. For $p_T^{\mathrm{miss}}$, the improvement is most evident in the dominant channels, while the tau categories remain broad, reflecting the intrinsic difficulty of the selected sample. The primary-lepton momentum error also tends to tighten under pre-training where that observable is defined, although the gain is less uniform than for $d_{\mathrm{PV}}$. Taken together with the classification results, Fig.~\ref{fig:classification_regression_results} confirms that pre-training does not merely sharpen the final classifier head: it improves the shared latent representation underlying both discrete and continuous physics targets.

\subsection{Attribution and representation diagnostics}

\begin{figure*}[htbp]
  \centering
  \includegraphics[width=0.95\textwidth]{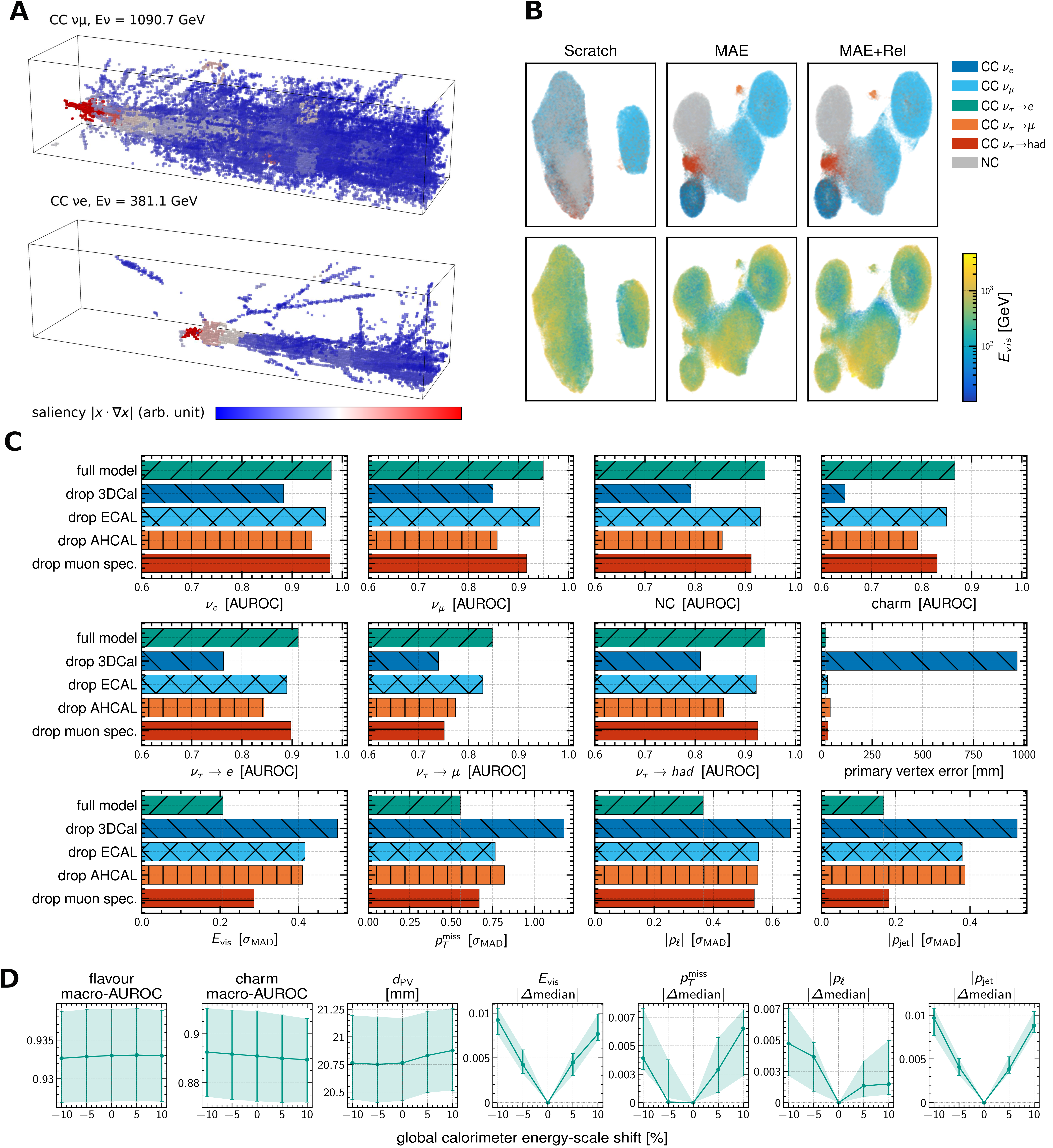}
  \caption{\textbf{Attribution and representation diagnostics for the learned representation.} \textbf{A} Two example patch-level 3DCal saliency maps for test charged-current events from the fine-tuned \maerelmodel{}. \textbf{B} Uniform manifold approximation and projection (UMAP) of the pooled latent representation for \scratchmodel{}, \maemodel{}, and \maerelmodel{}, coloured by true flavour (top) and visible energy, $E_{\mathrm{vis}}$ (bottom). \textbf{C} Detector-subsystem ablation for \maerelmodel{}, evaluated on the full test set after suppressing each detector branch at token fusion. Metrics are one-vs-rest flavour AUROC, charmed-quark macro-AUROC, mean three-dimensional vertex error, and $\sigma_{\mathrm{MAD}}$ for visible energy, missing transverse momentum, primary-lepton momentum and jet momentum. \textbf{D} Response of \maerelmodel{} to coherent global calorimeter energy-scale shifts applied at inference time in \threedcal{}, \ahcal{} and \ecal{}. The curves show flavour macro-AUROC, charm macro-AUROC, $d_{\mathrm{PV}}$, and the absolute change in the median signed fractional residual, $|\Delta\,\mathrm{median}[(x^{\mathrm{true}}-x^{\mathrm{reco}})/x^{\mathrm{true}}]|$, for $E_{\mathrm{vis}}$, $p_T^{\mathrm{miss}}$, $|p_\ell|$ and $|p_{\mathrm{jet}}|$, always measured relative to the nominal scale point. Shaded bands show paired bootstrap percentile intervals obtained by resampling the same evaluation events across all scale points. The panels should be read as diagnostic probes of model reliance and latent organisation rather than exhaustive explanations of the learned physics.}
  \label{fig:interpretability}
\end{figure*}

Figure~\ref{fig:interpretability} provides complementary evidence about what the fine-tuned encoder is using and how pre-training shapes the latent space. We treat these analyses as attribution and representation diagnostics, not as a complete physical interpretation of the learned model. They identify detector regions, detector branches and latent-space organisation associated with the predictions, while causal physics interpretation requires additional controlled studies. In panel~A, patch-level 3DCal saliency maps for two arbitrary test charged-current events show that attribution is concentrated near the interaction region and along a limited subset of downstream shower and track structures rather than being spread uniformly across all active patches. Even in the more crowded TeV-scale event, the highest-saliency regions follow the main event skeleton, suggesting that the model integrates localised interaction cues with extended downstream topology rather than relying on diffuse correlations alone. This pattern is consistent with the physics content of the tasks: flavour, charm and vertex inference depend both on the primary interaction region and on downstream track/shower development, especially for muonic and hadronic topologies.

Panel~B compares two-dimensional uniform manifold approximation and projection (UMAP) maps of the pooled latent representations for \scratchmodel{}, \maemodel{} and \maerelmodel{}, coloured by true flavour and visible energy. The \scratchmodel{} model already exhibits coarse organisation, while the pre-trained variants form cleaner low-dimensional structures, with better separated flavour groupings and smoother visible-energy progressions across the embedding. The coexistence of flavour clustering and energy ordering suggests that the representation organises events by multiple correlated physical attributes rather than along a single task-specific axis. Residual overlap, especially between neutral-current and $\nu_\mu$ CC events and among the tau channels, is consistent with the confusions seen in Fig.~\ref{fig:classification_regression_results}, indicating that pre-training sharpens but does not erase the hardest physical ambiguities. As a qualitative projection, UMAP is not itself a performance metric, but it is consistent with pre-training producing a more structured shared representation.

Panel~C tests how that representation uses the heterogeneous detector inputs. The result is clear: \threedcal{} provides the backbone of the event interpretation. Removing it lowers flavour macro-AUROC from 0.928 to 0.806 and macro-AUROC for charmed-quark classification from 0.866 to 0.647, while the mean vertex error rises from 21\,mm to 966\,mm. The auxiliary branches nevertheless provide targeted complementary information. Dropping \ahcal{} most strongly affects hadronic and neutral-current discrimination, reducing the one-vs-rest AUROC from 0.939 to 0.857 for $\nu_\tau$ CC $\rightarrow \mathrm{had}$ and from 0.939 to 0.854 for neutral-current events. Dropping the muon spectrometer has the clearest effect on muonic channels, reducing the one-vs-rest AUROC from 0.949 to 0.916 for $\nu_\mu$ CC and from 0.848 to 0.751 for $\nu_\tau$ CC $\rightarrow \mu$, and worsening the primary-lepton-momentum resolution. Removing \ecal{} degrades energy-related observables, with the visible-energy $\sigma_{\mathrm{MAD}}$ increasing from 0.207 to 0.417. These ablations should be read as showing which detector inputs the trained model uses for interactions whose vertices lie in \threedcal{}, not as an intrinsic ranking of subsystem importance. Because shower containment and leakage couple the downstream detector response, removing one detector branch from the network does not erase that subsystem's physical influence on the topology observed by the others. The ablation pattern therefore matches the detector design: \threedcal{} dominates, but the downstream subsystems contribute information in physically plausible, channel-dependent ways. This is detector attribution rather than full physics interpretability: it identifies model sensitivity to each input branch, but it does not prove that the network has learned a human-readable reconstruction algorithm.

Panel~D addresses a complementary question: whether the learned representation is fragile to coherent detector mismodelling. We apply a common global energy-scale shift to the calorimetric deposits in \threedcal{}, \ahcal{} and \ecal{} at inference time and reevaluate \maerelmodel{}, with paired bootstrap bands obtained by resampling the evaluated events across all scale points. In the current 50-batch scan, corresponding to 3200 events, flavour macro-AUROC remains essentially unchanged at 0.9326--0.9331, charm macro-AUROC varies from 0.8926 to 0.8894, and $d_{\mathrm{PV}}$ changes by only about 0.13\,mm across the full $\pm 10\%$ range. For the regression observables, we report the absolute change in the median signed fractional residual relative to the nominal scale point. This isolates the drift induced by the nuisance itself, rather than conflating it with any pre-existing calibration offset of the nominal model. Over the full scan, the largest observed drifts are 0.0092 for $E_{\mathrm{vis}}$, 0.0065 for $p_T^{\mathrm{miss}}$, 0.0047 for $|p_\ell|$, and 0.0097 for $|p_{\mathrm{jet}}|$. The scan is not a full detector-systematics programme, but it does suggest that the learned representation is not acutely brittle to an $\mathcal{O}(10\%)$ coherent calorimeter-scale bias. It should not be interpreted as covering generator, flux, channel-efficiency or subsystem-specific detector uncertainties; those require dedicated nuisance studies before deployment in an experimental analysis.

\subsection{Data efficiency study}

\begin{figure*}[htbp]
  \centering
  \includegraphics[width=0.95\textwidth]{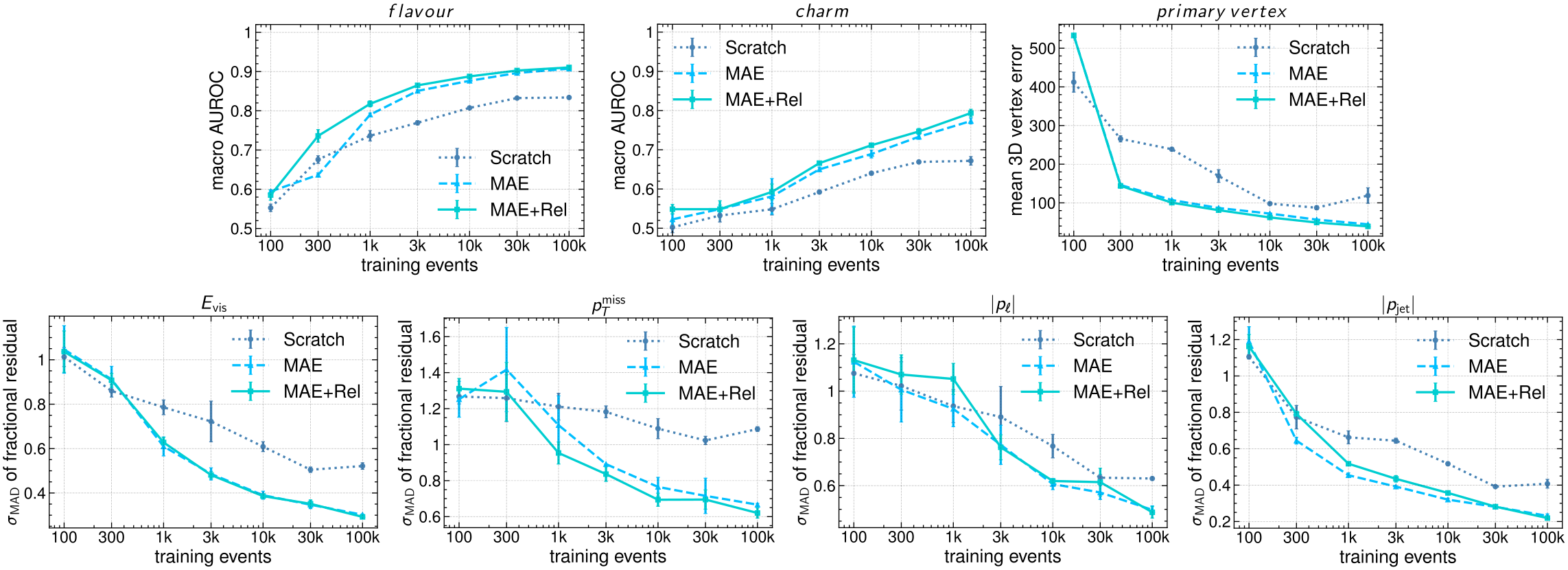}
  \caption{\textbf{Data-efficiency study for the three initialisation strategies.} Performance of \scratchmodel{}, \maemodel{} and \maerelmodel{} across labelled training budgets from 100 to 100,000 events. \textbf{A,B}, flavour and charm macro-AUROC. \textbf{C}, mean three-dimensional primary-vertex error. \textbf{D--G}, $\sigma_{\mathrm{MAD}}$ of the fractional residuals for visible energy, missing transverse momentum, primary-lepton momentum and jet momentum, respectively. Points are means across three random seeds and error bars show their standard deviation. Higher values are better for AUROC; lower values are better for regression and vertex errors.}
  \label{fig:data_eff}
\end{figure*}

Figure~\ref{fig:data_eff} isolates one of the most practically important effects of pre-training: reduced dependence on large labelled training sets. The study compares \scratchmodel{}, \maemodel{} and \maerelmodel{} across label budgets from $10^2$ to $10^5$ training events while keeping the validation and test sets fixed. Each point is averaged across three random seeds.

The completed three-way comparison shows that both pre-trained initialisations generally outperform \scratchmodel{}, with \maerelmodel{} giving the strongest overall performance across the study. Masked reconstruction accounts for a substantial part of the data-efficiency gain, while the relational continuation provides a further, task-dependent improvement. At $10^3$ labelled events, flavour macro-AUROC rises from 0.736 for \scratchmodel{} to 0.790 for \maemodel{} and 0.818 for \maerelmodel{}; the full pre-trained model already exceeds the 0.807 reached by \scratchmodel{} at $10^4$ events. Charm macro-AUROC at the same budget rises from 0.548 to 0.581 and 0.593, respectively, with \maerelmodel{} matching \scratchmodel{} trained on approximately three times more events. The classification advantage persists at $10^5$ events, where \maerelmodel{} remains the best-performing initialisation for both flavour and charm classification.

The regression results show the same broad ordering, although with more variation between targets and label budgets. At $10^3$ events, the mean vertex error falls from 239 mm for \scratchmodel{} to 107 mm for \maemodel{} and 100 mm for \maerelmodel{}. MAE+Rel also gives the clearest sustained advantage for missing transverse momentum and is generally strongest for vertex reconstruction. For jet momentum, however, $\sigma_{\mathrm{MAD}}$ at $10^3$ events falls from 0.663 to 0.454 and 0.517, so MAE alone is strongest at this budget; MAE-only is likewise competitive or superior at several budgets for visible-energy and primary-lepton-momentum regression. At the smallest budgets of 100--300 events, pre-training is not consistently advantageous across all continuous targets. Thus, the full relational objective is generally superior across the task suite, particularly for classification, vertexing and missing transverse momentum, but its advantage is not universal at every budget or for every regression target.

\subsection{\label{sec:robustness_transfer_results}Alternative-generator robustness and cross-domain transfer}

The broader motivation for pre-training is not only improved downstream performance within one simulated detector domain, but also the possibility of reusing representations across modelling choices, detector geometries, sensing modalities and energy regimes. We therefore collect the generator-stress and transfer results in a single summary table. Panel~A of Table~\ref{tab:robustness_transfer} tests generator dependence with a matched GENIE-to-NuWro comparison in the source detector. Panels~B and C test cross-domain transfer to a fine-grained plastic-scintillator benchmark and to PILArNet, respectively.

For the alternative-generator test, the same three fine-tuned checkpoints are evaluated without retraining on a NuWro~\cite{golan2012nuwro} sample matched one-to-one to the GENIE test sample at the level of the incoming neutrino, interaction vertex and target nucleus. The detector simulation and reconstruction inputs are kept in the same format as for the main evaluation. Further details on data generation are given in Sec.~\ref{sec:methods_nuwro}.

\begin{table*}[htb]
\centering
\caption{\textbf{Generator robustness and cross-domain transfer benchmarks.} A single table collects the three external-validation summaries. Panel A reports directly evaluated matched GENIE/NuWro performance in the source detector for all three models, with entries shown as GENIE/NuWro pairs. The classification columns are macro-averaged one-vs-rest AUROC values; the regression columns report $\sigma_{\mathrm{MAD}}$ of fractional residuals, except $d_{\mathrm{PV}}$, which is the median primary-vertex error. Panel B gives the full row-normalised confusion matrix for the fine-grained plastic-scintillator benchmark; published baselines are taken from Table 3 of Ref.~\cite{AlonsoMonsalve2023TrackFitting}, and rows sum to 1. Panel C gives the full PILArNet comparison; all published baseline rows are from Ref.~\cite{koh2023uncertainty}. The single-particle comparison marked with $^\ast$ is not strictly like-for-like because the reference paper reports the published values on a same-generation $1024^3$ dataset, whereas our rows use the public $768^3$-pixel release. The \maemodel{} rows in Panels B and C report the final MAE-only transfer results.}
\label{tab:robustness_transfer}
\footnotesize
\setlength{\tabcolsep}{2.6pt}
\renewcommand{\arraystretch}{0.94}
\makebox[\textwidth][c]{%
\begin{tabular}{lccccccc}
\multicolumn{8}{l}{\textbf{A. Matched GENIE-to-NuWro stress test}} \\
\toprule
Model & Flav. AUC & Charm AUC & $E_{\mathrm{vis}}$ & $p_T^{\mathrm{miss}}$ & $|p_\ell|$ & $|p_{\mathrm{jet}}|$ & $d_{\mathrm{PV}}$ [mm] \\
\midrule
\scratchmodel{} & 0.893/0.889 & 0.790/0.647 & 0.386/0.390 & 0.705/0.721 & 0.447/0.448 & 0.264/0.274 & 22.1/22.3 \\
\maemodel{} & 0.929/0.925 & 0.863/0.720 & 0.256/0.261 & 0.535/0.552 & 0.395/0.397 & 0.190/0.199 & 16.9/17.1 \\
\maerelmodel{} & 0.932/0.928 & 0.864/0.721 & 0.254/0.258 & 0.543/0.552 & 0.396/0.398 & 0.187/0.196 & 16.1/16.3 \\
\bottomrule
\end{tabular}
}

\vspace{0.35em}
\makebox[\textwidth][c]{%
\begin{tabular}{c r c c c c}
\multicolumn{6}{l}{\textbf{B. Transfer to the fine-grained plastic-scintillator benchmark}} \\
\toprule
\multicolumn{1}{c}{True class} &
\multicolumn{1}{c}{Method} &
\multicolumn{1}{c}{Pred.\ $p$} &
\multicolumn{1}{c}{Pred.\ $\pi^\pm$} &
\multicolumn{1}{c}{Pred.\ $\mu^\pm$} &
\multicolumn{1}{c}{Pred.\ $e^\pm$} \\
\midrule
\multirow{6}{*}{$p$}
& GBDT-Transformer~\cite{AlonsoMonsalve2023TrackFitting} & 0.907 & 0.067 & 0.007 & 0.019 \\
& GBDT-RNN~\cite{AlonsoMonsalve2023TrackFitting} & 0.896 & 0.073 & 0.006 & 0.025 \\
& GBDT-SIR-PF~\cite{AlonsoMonsalve2023TrackFitting} & 0.891 & 0.077 & 0.008 & 0.024 \\
& \scratchmodel{} (ours) & 0.919 & 0.060 & 0.003 & 0.018 \\
& \maemodel{} (ours) & 0.934 & 0.049 & 0.005 & 0.012 \\
& \maerelmodel{} (ours) & \textbf{0.943} & 0.040 & 0.006 & 0.011 \\
\midrule
\multirow{6}{*}{$\pi^\pm$}
& GBDT-Transformer~\cite{AlonsoMonsalve2023TrackFitting} & 0.057 & \textbf{0.643} & 0.041 & 0.259 \\
& GBDT-RNN~\cite{AlonsoMonsalve2023TrackFitting} & 0.080 & 0.623 & 0.036 & 0.261 \\
& GBDT-SIR-PF~\cite{AlonsoMonsalve2023TrackFitting} & 0.080 & 0.606 & 0.042 & 0.272 \\
& \scratchmodel{} (ours) & 0.062 & 0.532 & 0.267 & 0.139 \\
& \maemodel{} (ours) & 0.056 & 0.584 & 0.244 & 0.116 \\
& \maerelmodel{} (ours) & 0.053 & 0.609 & 0.227 & 0.111 \\
\midrule
\multirow{6}{*}{$\mu^\pm$}
& GBDT-Transformer~\cite{AlonsoMonsalve2023TrackFitting} & 0.071 & 0.190 & 0.595 & 0.144 \\
& GBDT-RNN~\cite{AlonsoMonsalve2023TrackFitting} & 0.089 & 0.233 & 0.506 & 0.172 \\
& GBDT-SIR-PF~\cite{AlonsoMonsalve2023TrackFitting} & 0.126 & 0.236 & 0.517 & 0.121 \\
& \scratchmodel{} (ours) & 0.020 & 0.173 & 0.661 & 0.146 \\
& \maemodel{} (ours) & 0.020 & 0.108 & 0.723 & 0.149 \\
& \maerelmodel{} (ours) & 0.019 & 0.079 & \textbf{0.748} & 0.155 \\
\midrule
\multirow{6}{*}{$e^\pm$}
& GBDT-Transformer~\cite{AlonsoMonsalve2023TrackFitting} & 0.020 & 0.199 & 0.009 & 0.772 \\
& GBDT-RNN~\cite{AlonsoMonsalve2023TrackFitting} & 0.027 & 0.200 & 0.007 & 0.766 \\
& GBDT-SIR-PF~\cite{AlonsoMonsalve2023TrackFitting} & 0.017 & 0.237 & 0.006 & 0.740 \\
& \scratchmodel{} (ours) & 0.019 & 0.179 & 0.091 & 0.712 \\
& \maemodel{} (ours) & 0.015 & 0.108 & 0.109 & 0.768 \\
& \maerelmodel{} (ours) & 0.012 & 0.084 & 0.117 & \textbf{0.787} \\
\bottomrule
\end{tabular}
}

\vspace{0.35em}
\makebox[\textwidth][c]{%
\begin{tabular}{rcc@{\hspace{0.8em}}cc}
\multicolumn{5}{l}{\textbf{C. Transfer to PILArNet particle-identification benchmarks}} \\
\toprule
& \multicolumn{2}{c}{Single-particle classification$^\ast$}
& \multicolumn{2}{c}{Multi-particle classification} \\
\cmidrule(r{0.45em}){2-3}
\cmidrule(l{0.45em}){4-5}
\multicolumn{1}{c}{Method}
& \metric{Accuracy} & \metric{AUROC}
& \metric{Accuracy} & \metric{AUROC} \\
\midrule
Deterministic~\cite{koh2023uncertainty}      & \metric{0.8656} & \metric{0.753} & \metric{0.9604} & \metric{0.938} \\
Naive Ensembles~\cite{koh2023uncertainty}     & \metric{0.8844} & \metric{0.827} & \metric{0.9640} & \metric{0.944} \\
Bootstrap Ensembles~\cite{koh2023uncertainty} & \metric{0.9014} & \metric{0.842} & \metric{0.9644} & \metric{0.942} \\
MC Dropout~\cite{koh2023uncertainty}          & \metric{0.8734} & \metric{0.795} & \metric{--} & \metric{--} \\
EDL-MLL~\cite{koh2023uncertainty}             & \metric{0.8622} & \metric{0.762} & \metric{0.9604} & \metric{0.935} \\
EDL-BR~\cite{koh2023uncertainty}              & \metric{0.8253} & \metric{0.701} & \metric{0.9223} & \metric{0.900} \\
EDL-Brier~\cite{koh2023uncertainty}           & \metric{0.8751} & \metric{0.748} & \metric{0.9596} & \metric{0.911} \\
\scratchmodel{} (ours)                        & \metric{0.8798} & \metric{0.834} & \metric{0.9333} & \metric{0.922} \\
\maemodel{} (ours)                      & \metric{0.9050} & \metric{0.875} & \metric{0.9580} & \metric{0.944} \\
\maerelmodel{} (ours)                         & \metric{\textbf{0.9154}} & \metric{\textbf{0.891}} & \metric{\textbf{0.9662}} & \metric{\textbf{0.951}} \\
\bottomrule
\end{tabular}
}
\end{table*}

Panel~A gives a deliberately mixed, and therefore useful, outcome. The directly evaluated GENIE results preserve the in-domain ordering: \scratchmodel{} is weakest overall, \maemodel{} recovers most of the flavour and kinematic performance, and \maerelmodel{} is marginally strongest on the classification metrics and on most reported regression quantities. The flavour macro-AUROC values are 0.893, 0.929 and 0.932 for \scratchmodel{}, \maemodel{} and \maerelmodel{}, respectively; the corresponding charm macro-AUROCs are 0.790, 0.863 and 0.864. All three NuWro rows are direct evaluations of the corresponding fine-tuned checkpoints. For \maerelmodel{}, flavour macro-AUROC changes only mildly from 0.932 to 0.928, and the kinematic resolutions and median vertex error likewise show small changes. By contrast, charm macro-AUROC decreases from 0.864 to 0.721. The underlying one-vs-rest charm AUCs for no-charm, charm$\to e$, charm$\to\mu$ and charm$\to\mathrm{had}$ change from $(0.877,0.809,0.891,0.877)$ to $(0.727,0.681,0.755,0.719)$. Taken together, the small measured shifts in flavour identification, kinematic reconstruction and vertex localisation are encouraging within this matched generator comparison, but do not establish generator independence. Charm tagging shows substantial sensitivity to generator-dependent heavy-flavour production, hadronisation and event topology.

Panels~B and C test whether the source-domain encoder carries useful information outside the original detector, and also clarify the role of MAE-only pre-training in transfer. The scintillator benchmark of Ref.~\cite{AlonsoMonsalve2023TrackFitting} is technologically close to \threedcal{}, because both detectors use fine-grained plastic scintillator, but differs in detector dimensions, magnetic environment, task definition and energy scale. Relative to target-domain training from random initialisation, the \maemodel{} row already improves the diagonal entries for all four classes: from 0.919 to 0.934 for protons, from 0.532 to 0.584 for charged pions, from 0.661 to 0.723 for muons and from 0.712 to 0.768 for electrons. The full \maerelmodel{} then increases them further to 0.943, 0.609, 0.748 and 0.787. This indicates that masked reconstruction accounts for a large fraction of the transferred geometric and calorimetric prior, while the relational objective provides an additional gain, especially for the pion, muon and electron classes.

PILArNet~\cite{adams2020pilarnet} introduces a stronger shift, from scintillating voxels to liquid-argon time-projection-chamber data and from neutrino-event analysis to particle-level classification. In the single-particle task, the \maemodel{} result improves over \scratchmodel{} from 0.8798 to 0.9050 in accuracy and from 0.834 to 0.875 in entropy-based AUROC; \maerelmodel{} further reaches 0.9154 and 0.891. The single-particle comparison with published values should be read carefully because of the public-resolution caveat in Table~\ref{tab:robustness_transfer}. The multi-particle task is directly comparable on the public PILArNet benchmark: the \maemodel{} row improves over \scratchmodel{} from 0.9333 to 0.9580 in accuracy and from 0.922 to 0.944 in AUROC, while \maerelmodel{} reaches 0.9662 and 0.951, edging past the best published values in both metrics. Taken together, the two transfer studies suggest that masked reconstruction already learns reusable detector geometry, and that the relational objectives further sharpen the representation when the target task demands particle-level or multi-particle discrimination.

\section{\label{sec:discussion}Discussion}

The main finding of this study is that self-supervised pre-training becomes most valuable precisely where energy-frontier neutrino events are hardest to interpret. A single pre-trained encoder improves flavour classification, charmed-quark identification, kinematic regression and vertex reconstruction, and the gains are largest in channels where shower overlap, secondary activity and partial containment make the event topology genuinely ambiguous. This matters because these channels carry direct physics significance: they are closely tied to the tau-neutrino and heavy-flavour measurements that motivate forward collider-neutrino programmes. In this regime, representation learning is not simply refining an already mature reconstruction chain, but it is helping to establish a viable analysis strategy.

The comparison between \maemodel{} and \maerelmodel{} also clarifies what the pre-training objective is contributing. Masked reconstruction alone already captures a substantial fraction of the global shower geometry and cross-detector context, as reflected in the improvements for the dominant flavour categories and in the broadly better regression behaviour with respect to \scratchmodel{}. The relational targets add a more local and semantic constraint, and their benefit is most visible in the lower-yield tau and charmed-quark channels, where ghost suppression, secondary structure and local ambiguities matter most. This interpretation is now supported by three-way comparisons in the in-domain, data-efficiency and cross-domain transfer studies. Across label budgets, MAE alone accounts for most of the improvement over \scratchmodel{}, while MAE+Rel provides a consistent additional classification gain and selected improvements, notably for missing transverse momentum and vertex reconstruction, without uniformly improving every regression target. Recent work in detector and neutrino machine learning has similarly shown that self-supervised objectives can improve robustness or transferability through contrastive learning, masked point modelling and related pretext tasks~\cite{Wilkinson2025Contrastive,Young2026Particle,Yu:2025t0}. Related studies in broader particle physics have reported transferable gains from self-supervised pre-training and fine-tuning, although mostly in collider-jet or analysis-level settings rather than heterogeneous detector reconstruction~\cite{Harris2025RS3L,Heinrich2024MPM,Birk2024OmniJet,Vigl2024Finetuning}. What distinguishes the present setting is the combination of severe event complexity, heterogeneous detector inputs and joint downstream inference. The present results therefore suggest that, for dense neutrino events, reconstruction-style objectives are a strong starting point but not the whole answer: physics-aware local supervision can further sharpen the latent representation where the ambiguity is greatest.

The attribution and representation diagnostics are consistent with that picture. The saliency examples emphasise the interaction region and the main downstream structures rather than diffuse occupancy alone. The latent-space UMAP maps show a more orderly geometry under \maerelmodel{}, with clearer flavour grouping and smoother visible-energy variation than \scratchmodel{}. Detector-branch ablations then recover a physically sensible division of labour: \threedcal{} carries the backbone of the inference, \ahcal{} contributes most strongly to hadronic and neutral-current discrimination, the muon spectrometer to muonic signatures, and \ecal{} to energy-related observables. A complementary coherent energy-scale stress test produces only mild drifts over a $\pm 10\%$ scan, with nuisance-induced bias changes at the level of a few $10^{-3}$ to $10^{-2}$ in the current evaluation, suggesting that the representation is not overly sensitive to global calorimetric miscalibration. These analyses do not by themselves explain the model exhaustively, but they strengthen the case that the learned representation is structured rather than ad hoc. They should therefore be viewed as physically motivated consistency checks, not as proof that the network has learned a fully interpretable event-reconstruction procedure.

The data-efficiency study strengthens that interpretation. Other recent detector studies have reported that self-supervised pre-training can reduce downstream label requirements~\cite{Vigl2024Finetuning,Young2026Particle,babicz2026transformerbasedpulseshapediscrimination}, but those results were obtained in simpler settings, such as individual particle trajectories or detector pulse waveforms. Here the effect is large enough to be operationally meaningful in a heterogeneous, multi-task neutrino problem. With roughly $10^3$ labelled events, \maerelmodel{} reaches flavour-classification performance exceeding \scratchmodel{} trained on roughly $10^4$ events, while \maemodel{} already provides most of the improvement in jet regression and vertex reconstruction. The separation of the two pre-trained variants shows that the reduction in label requirements is principally a masked-reconstruction effect, supplemented by relational gains that are strongest for classification and selected continuous targets. That leftward shift of the performance frontier matters because, in an energy-frontier neutrino programme, labels are not a cheap by-product. Less abundant channels require dedicated simulation campaigns, systematic variations are costly, and many target quantities depend on expensive truth association or high-level reconstruction. A representation that reduces the demand for supervision therefore changes not only how efficiently models can be trained, but also which physics studies are practical to pursue.

The transfer results place the work in a broader context. Transfer learning in neutrino physics has so far been used mainly to adapt image-based backbones to downstream classification tasks, or to reduce domain mismatch between nearby source and target problems~\cite{Chappell2022Application,Babicz2022Adversarial,Bonilla2026Transfer}. Representation-learning studies have also begun to assess whether pre-trained encoders can improve robustness within a detector family~\cite{Wilkinson2025Contrastive}, while broader particle-physics work has started to test transfer from pre-trained encoders across tasks and datasets in collider settings~\cite{Birk2024OmniJet,Vigl2024Finetuning}. The transfer experiments here are deliberately staged. We first move to a public scintillator detector that is close to \threedcal{} in sensing modality and voxel size but different energy regime, then to the public PILArNet benchmark, which introduces a different detector technology and a different energy regime too. In the scintillator case, transfer improves the class-conditional confusion-matrix diagonals for all four particle species and surpasses the strongest published baseline for protons, muons and electrons while approaching it for pions. In PILArNet, a source encoder pre-trained on TeV-scale neutrino interactions adapts to a public LArTPC particle-identification benchmark, improves both single-particle and multi-particle performance over scratch training, and on the multi-particle task also edges past the strongest published ensemble baseline~\cite{koh2023uncertainty}. Taken together, these results suggest that the pre-training captures structure that is useful well beyond the original detector, task and energy range.

The conclusions should be read with clear limits in mind. The study is simulation-based, the pre-training evidence is mainly qualitative, and the relational targets rely on simulation truth. Those targets are useful precisely because they encode local detector semantics, but they may also inherit generator and detector-model assumptions. The matched NuWro evaluation in Sec.~\ref{sec:robustness_transfer_results} adds a controlled check of this risk. The close agreement in flavour identification, kinematic reconstruction and vertex localisation provides encouraging evidence that the learned representation is robust to generator-specific differences for these tasks within the matched configuration. Charm tagging sets an important boundary: its substantial degradation indicates sensitivity to neutrino-interaction and heavy-flavour production, hadronisation and topology. We therefore interpret the NuWro comparison as evidence of task-dependent generator robustness, not generator independence or complete systematic validation. Broader generator reweighting, flux variations, detector-response mismodelling and experimental-data calibration remain necessary before deployment. Recent work on masked modelling and domain adaptation in neutrino experiments has highlighted the importance of testing whether learned representations remain stable under simulation mismodelling and domain shift~\cite{Babicz2022Adversarial,Yu:2025t0,Bonilla2026Transfer}. Stronger claims about broadly reusable detector models will require wider transfer studies, dedicated stress tests and, ultimately, validation on experimental data.

A related deployment question is how errors in one detector branch propagate through a shared multimodal latent representation. The ablation study shows that the branches contribute in channel-dependent ways, but suppressing a branch is not the same as modelling a biased or inefficient branch. A realistic deployment programme should therefore include subsystem-specific perturbations, such as independent calorimeter-scale shifts, dead or inefficient channels, altered thresholds and noise, muon-spectrometer inefficiencies, and detector-response mismodelling, and should examine whether such perturbations bias all downstream tasks through the fused representation. Uncertainty quantification is also essential for experimental use. Possible extensions include nuisance-aware training, calibrated ensembles, Bayesian or evidential task heads, conformal prediction, and latent-space or reconstruction-based out-of-distribution scores. Such tools could make the learned representation useful not only for supervised reconstruction, but also for rare-event triage or discovery-oriented searches. The present work does not demonstrate anomaly detection or unseen-physics discovery; it instead establishes the representation-learning basis on which such studies could be built.

We therefore do not claim that a general detector foundation model has been achieved. Rather, the results show that several ingredients usually discussed separately can coexist in one detector-aware sparse encoder: self-supervised pre-training, joint downstream fine-tuning, meaningful low-label gains, and non-trivial transfer beyond the source detector. For energy-frontier neutrino physics, where event complexity makes artificial intelligence a requirement rather than an optional convenience, that is already a substantive result. It also outlines a concrete path towards more general detector encoders and motivates further work on hybrid pre-training objectives and domain adaptation.

\section{\label{sec:method}Methods}

\subsection{\label{sec:case_study}Case-study detector and simulated data}

We use the proposed \fasercal{} concept at the Large Hadron Collider at CERN as a representative energy-frontier case study~\cite{CERN-FASER-NOTE-2026-004}. In this setup, neutrino interactions occur inside the highly granular \threedcal{}, a 10-module detector with $48\times48\times200$ scintillator voxels in total, which is followed downstream by the electromagnetic calorimeter (\ecal{}), the hadronic calorimeter (\ahcal{}), and a muon spectrometer. We refer to the primary calorimeter as \threedcal{} throughout, reserving \fasercal{} for the broader detector concept. An example event spanning the detector subsystems is shown in Supplementary Figure~\ref{fig:event_display_antinu_mu_cc}.

Run-4 forward neutrino fluxes from light-hadron and charm-hadron sources are used to generate interactions with GENIE v3.04.00~\cite{andreopoulos2010genie}, tau-lepton and charm-hadron decays are modelled with PYTHIA8~\cite{bierlich2022pythia8}, and the resulting particles are propagated through the full detector geometry with Geant4~\cite{agostinelli2003geant4,CERN-FASER-NOTE-2026-004}. This matters physically because the flavour composition and energy spectrum depend on the parent-hadron origin: charm dominates the $\nu_\tau$ component and contributes substantially to the high-energy $\nu_e$ flux in the far-forward region. The collaboration simulation and reconstruction framework underlying these studies is publicly available at \url{https://github.com/rubbiaa/FASER}.

The nominal simulated sample corresponds to 101~ab$^{-1}$ of integrated luminosity and yields 1,118,058 neutrino interactions in the \threedcal{} modules. To improve training statistics for the less abundant and most topologically difficult channels, we add 108,317 dedicated $\nu_\tau$ charged-current interactions from an enriched sample corresponding to 3,700~ab$^{-1}$. The combined event pool is split once into train, validation and test partitions of 85/5/10. Because this pooled set over-represents $\nu_\tau$ events, all the results presented in this manuscript are reweighted at evaluation time so the $\nu_\tau$ abundance matches the nominal unbiased sample; this is the procedure implemented in the analysis code used to generate the paper figures.

\begin{figure*}[htpb]
  \centering
  \includegraphics[width=1.0\linewidth]{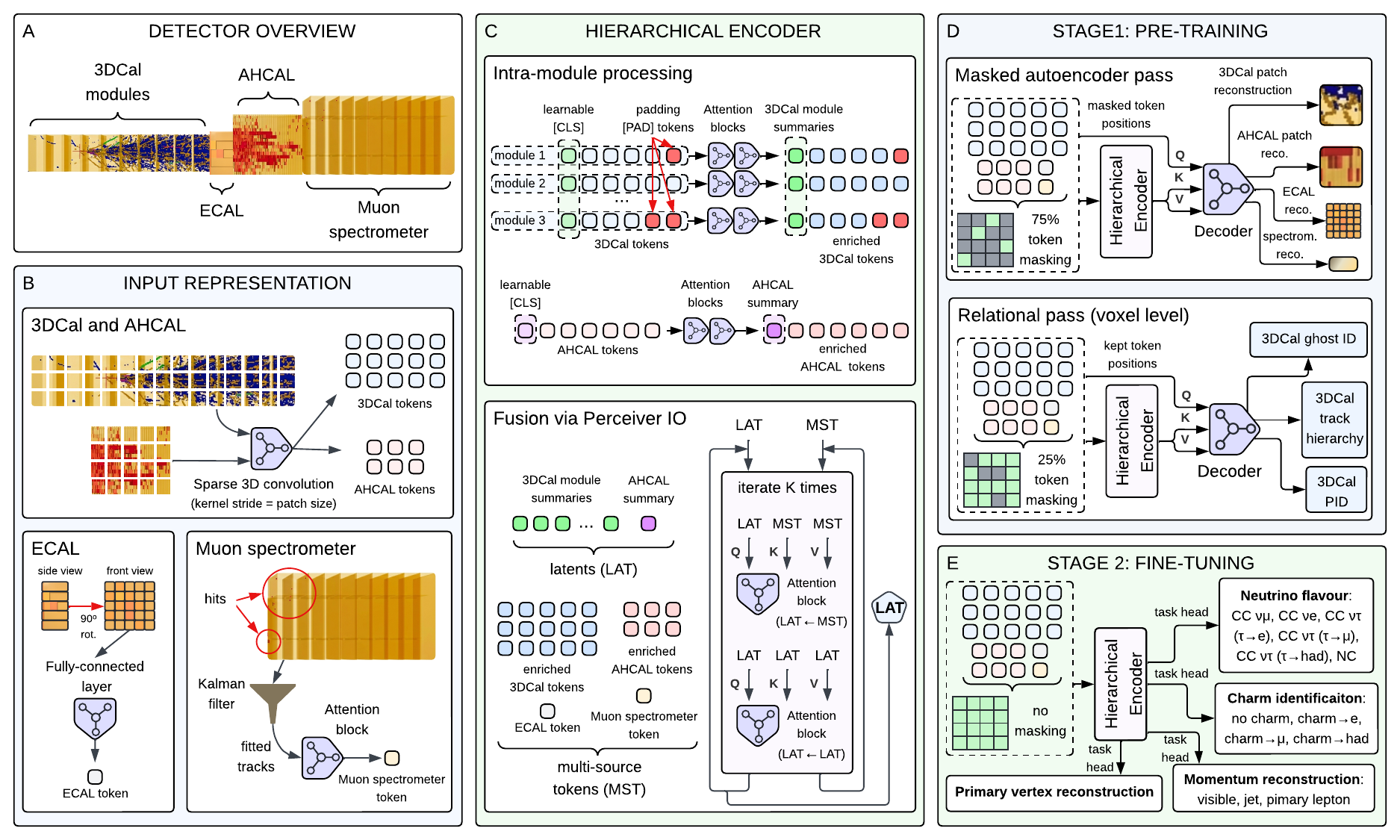}
  \caption{\textbf{Overview of the framework.} \textbf{A} \fasercal{} case study, with the \threedcal{} as the primary interaction volume followed by the \ecal{}, \ahcal{} and muon spectrometer. \textbf{B} The sparse \threedcal{} and \ahcal{} voxel inputs, dense $5\times5$ \ecal{} energy matrix and variable-length muon-spectrometer summaries are converted into detector-specific tokens. \textbf{C} A hierarchical encoder performs module-level self-attention before Perceiver-IO fusion of the heterogeneous detector streams. \textbf{D} Stage 1 pre-training combines masked reconstruction with a relational voxel-level pass. \textbf{E} The pre-trained encoder is then fine-tuned jointly on event-level classification and regression tasks. A representative event spanning the four input subsystems is shown in Supplementary Figure~\ref{fig:event_display_antinu_mu_cc}.}
  \label{fig:model}
\end{figure*}

The detector inputs are heterogeneous by construction. The \threedcal{} provides sparse three-dimensional voxel hits with charge information and simulation-derived voxel labels used only during pre-training. The \ahcal{} provides a second sparse calorimetric volume at coarser granularity. The \ecal{} is represented as a compact $5\times5$ energy matrix, and the muon spectrometer provides up to ten hit-measuring planes per track. This mixture of sparse volumetric inputs, dense global summaries and variable-length track information is precisely why a unified encoder is non-trivial. Before tokenisation, the \threedcal{} and \ahcal{} inputs are treated as sparse lists of occupied voxel coordinates and charge-like readout values, with charge transformed using the same logarithmic preprocessing used throughout training. The \ecal{} input is a dense two-dimensional energy summary, encoded as a single detector token after a small learned projection. The muon-spectrometer input is variable length: track or hit-plane summaries are embedded and pooled into a compact spectrometer token, together with presence/count information. Fig.~\ref{fig:model}B shows how these four native detector representations are converted into detector-specific tokens and fused, while Supplementary Figure~\ref{fig:event_display_antinu_mu_cc} provides a representative event spanning the same four subsystems. Read together, the two figures connect the physical detector response to the precise inputs and preprocessing supplied to the model before multimodal fusion.

Simulation also provides auxiliary truth that is useful during pre-training. In addition to event-level labels for the downstream tasks, we exploit voxel-level occupancy, charge, ghost labels, hierarchy labels and particle-category labels inside the \threedcal{}. Ghost labels identify reconstructed voxels with no matched true particle. Hierarchy labels separate background activity from voxels dominated by primary or secondary particles, while particle-category labels group matched truth deposits into electromagnetic, muonic and hadronic activity. The ghost target is binary, but the two semantic targets are not hard one-hot labels: a reconstructed voxel can be matched to several true deposits, so hierarchy and particle-category supervision are represented as soft per-voxel distributions obtained by summing the matched fractional contribution weights for each class and normalising within the voxel. This formulation is necessary because dense shower regions often mix contributions from several particles within a single reconstructed voxel.

\subsection{\label{sec:methods_nuwro}Matched alternative-generator sample}

To probe the dependence of the trained models on the neutrino-interaction generator, we constructed a NuWro~\cite{golan2012nuwro} sample matched event-by-event to the GENIE test set. For each GENIE test interaction we generated a single NuWro interaction with the same incoming-neutrino four-momentum (energy, flavour and direction), the same struck target nucleus, and the same current (charged or neutral); the primary-interaction vertex was fixed to the GENIE value. Given this initial state, NuWro selected the exclusive reaction channel and produced the hadronic final state according to its own cross-section, nuclear and hadronisation models, with deep-inelastic hadronisation handled by PYTHIA6~\cite{sjostrand2006pythia6}. The matching holds the flux, detector geometry and target composition fixed, making the comparison primarily sensitive to differences in the primary-interaction and hadronisation models rather than to exposure or acceptance.

Tau leptons and charmed hadrons were propagated undecayed out of the primary generator and decayed by a common downstream stage based on PYTHIA8~\cite{bierlich2022pythia8}, identical to the one used in the nominal GENIE production. The $\tau$ and heavy-flavour decay modelling is therefore shared between the two samples, and the $\tau\!\to\! e/\mu/\mathrm{had}$ and charm$\,\to e/\mu/\mathrm{had}$ categories are defined from the decay products exactly as in Sec.~\ref{sec:classification_results}. Charm production in NuWro proceeds through the slow-rescaling mechanism; the resulting charmed charged-current fraction is $9.2\%$, compared with $9.66\%$ in the GENIE sample, and the simulated $\tau$-decay branching fractions are consistent with their Standard-Model values. 
Because the $\tau$ leptons and charmed hadrons produced by the two generators carry different kinematics, their decays are independent realisations of a common decay model rather than event-by-event replicas; only the incoming neutrino state, vertex and target nucleus are matched.

Neutrino--electron elastic-scattering events have no nuclear-target counterpart in this matched configuration and were not reproduced; they constitute a negligible fraction of the sample.

To preserve $\nu_\tau$ statistics, the matching reproduces the full composition of the GENIE test set, including its $\tau$-enhanced beam component. Each matched event was passed through the same \fasercal{} Geant4 detector simulation and reconstruction chain as the nominal sample, producing identical input formats. The final evaluated NuWro sample contains 120,611 events.

\subsection{\label{sec:architecture}Sparse multimodal encoder}

All headline experiments use the same base encoder. The model follows the design sketched in Fig.~\ref{fig:model}. Sparse 3D convolutions use the SpConv framework~\cite{spconv2020} to convert the \threedcal{} and \ahcal{} voxel grids into patch tokens while operating only on occupied regions~\cite{graham2018submanifold}. The \threedcal{} is tokenised in patches of $12 \times 12 \times 10$ voxels, yielding a $4 \times 4 \times 20$ patch grid of up to 320 tokens, of which only those overlapping at least one active voxel are retained. The \ahcal{} is tokenised in patches of $6 \times 6 \times 5$ voxels ($3 \times 3 \times 8$ grid, up to 72 tokens). This sparse tokenisation makes the computational cost scale with detector occupancy rather than with the total instrumented volume.

The token sequence is then processed in two stages. First, \threedcal{} tokens are grouped by detector module (the detector comprises ten longitudinal modules, each spanning two patch planes in depth) and processed through module-level self-attention blocks augmented with learned module-position embeddings and per-module class tokens. This structure allows the model to capture local shower patterns within each module before mixing information across the detector. The \ahcal{} branch is processed through a parallel self-attention stack with its own class tokens. All attention layers use an embedding dimension of 384 with 12 heads (head dimension of 32) and an MLP ratio of 4.

Second, a Perceiver-IO bottleneck~\cite{jaegle2021perceiver} fuses the calorimetric tokens with compact representations of the \ecal{} (encoded as a single token from its $5 \times 5$ energy matrix) and muon spectrometer (encoded from up to ten hit-measuring planes per track). Learned detector-type embeddings distinguish the sources. The cross-attention and self-attention layers in the Perceiver loop produce a fixed-size latent representation that can be consumed by either the pre-training decoder or the downstream task heads.

The downstream fine-tuning and inference model used in the headline experiments contains 31,936,341 trainable model parameters; including the six learned homoscedastic loss-weight scalars used only during optimisation gives 31,936,347 optimised scalars during fine-tuning. This hierarchy is important for two reasons. Computationally, it avoids global attention over an unnecessarily large token sequence. Scientifically, it respects the detector's physical organisation while still allowing global event context to emerge in the latent space. The architecture therefore provides a natural compromise between local geometric fidelity and cross-detector integration. When loading pre-trained weights into the fine-tuning model, matching layers are transferred and the additional layers are randomly initialised. Full architectural specifications are provided in Supplementary Information~\ref{app:architecture}.

\subsection{\label{sec:pretraining_method}Stage 1: self-supervised pre-training}

Pre-training combines two complementary objectives applied in a two-phase schedule. In both phases, the core objective is masked reconstruction in the spirit of MAE~\cite{he2022masked}: 75\% of occupied calorimeter patches are randomly masked, and a lightweight decoder (embedding dimension 256, 8 heads) cross-attends from mask-token queries to the encoder's latent representation to predict voxel-level occupancy and charge in the missing regions. To map a single decoder token back to the many voxels within its patch, a multi-rank separable basis (initialised from discrete cosine transform (DCT) coefficients) is used. Auxiliary detector inputs (\ecal{} and muon spectrometer) are also masked and reconstructed when applicable. The masking is occupancy-aware, so the model concentrates capacity on non-trivial targets.

In the first phase, the model is trained for 400 epochs using masked reconstruction alone, allowing the encoder to learn broad spatial correlations and cross-detector context. The checkpoint at the end of this phase defines the \maemodel{} encoder used in the downstream comparisons. In the second phase, training continues from this checkpoint for 100 additional epochs while introducing, with probability 0.5 per batch, a relational forward pass in which the encoder predicts voxel-level ghost labels, hierarchy labels and particle-category labels on kept \threedcal{} patches (with a lower mask ratio of 0.25). Ghost remains a hard binary target, whereas hierarchy and particle-category supervision use soft labels built from the normalised fractional contributions of all matched truth deposits in a voxel. The semantic classes are therefore allowed to overlap at voxel level, which makes the task particularly demanding in dense shower regions. The checkpoint at the end of the second phase defines the \maerelmodel{} encoder. The two-phase schedule avoids premature saturation on the relational tasks, which are inherently easier than masked reconstruction. The MAE+Rel continuation required 427 GPU-hours in addition to the 584 GPU-hours used for MAE pre-training, an incremental pre-training cost of 73\%. The relational decoder and heads are discarded before fine-tuning, so this continuation does not change the downstream parameter count or inference path. The incremental gains over MAE alone are modest for many aggregate AUROC values but are larger for selected low-yield operating points; for example, the maximum figure of merit for $\nu_\tau$ CC$\to e$ increases from 1.03 to 1.35. The full objective therefore trades substantial additional one-off pre-training cost for targeted gains in several difficult channels, rather than a uniform improvement across all observables.

All pre-training losses are combined using learned homoscedastic uncertainty weights~\cite{kendall2018multi}. The reconstruction losses use a hybrid formulation that includes soft-chamfer style~\cite{fan2017pointset} and distance-weighted regression terms, so that predictions close to the true sparse support are penalised more gently than those that miss the relevant structure entirely. The optimiser is AdamW~\cite{loshchilov2019adamw} with a base learning rate of $10^{-4}$ (linearly scaled by effective batch size), $\beta_1 = 0.9$, $\beta_2 = 0.95$, and weight decay $0.05$. Phase~1 uses 40 warm-up epochs and cosine annealing over 360 epochs; the shorter Phase~2 continuation uses 10 warm-up epochs and cosine annealing over 90 epochs. Pre-training is distributed across two nodes of four NVIDIA GH200 GPUs each (eight GPUs total) with a per-GPU batch size of 512. The active training cost is summarised in Supplementary Information~\ref{app:model_scale}.

\subsection{\label{sec:finetuning_method}Stage 2: joint fine-tuning}

For downstream learning, the MAE decoder is discarded and the shared encoder is retained. Task-specific tokens or lightweight heads read out the latent representation and jointly predict neutrino flavour, charmed-quark category, visible momentum, jet momentum, and the primary interaction vertex. Primary-lepton momentum is derived from the reconstructed event components and is used in the evaluation plots. For a fair comparison, all downstream variants use the same fine-tuning architecture and the same random initialisation for task-specific heads; only the encoder initialisation differs between \scratchmodel{}, \maemodel{} and \maerelmodel{}.

Fine-tuning is multi-task throughout. This matters because the downstream observables are physically coupled: flavour identification, charm production, visible energy flow and vertex position all depend on a common interpretation of the same event. Joint fine-tuning therefore tests whether the pre-trained representation is useful as a shared basis for multiple measurements rather than for a single optimised classifier.

\subsection{\label{sec:transfer_method}Transfer-learning targets and adaptation}

Transfer experiments retain the transferable core of the source encoder while replacing detector-specific components. In practice, the attention blocks, latent cross-attention blocks, latent self-attention blocks, normalisation layers and global query token are inherited when shapes and semantics match, whereas detector-specific patch embeddings, positional encodings, detector-branch embeddings and task heads are reinitialised for the target domain. Scratch baselines use the same target architectures without transferred weights.

One target is the public fine-grained plastic-scintillator benchmark associated with Ref.~\cite{AlonsoMonsalve2023TrackFitting}. This dataset provides four-class particle identification on isolated charged-particle tracks at GeV scale energies in a three-dimensional detector built from $1\,\mathrm{cm}^3$ scintillator elements. We train on the provided public training split, reserve 5\% of it for validation, and evaluate on the provided testing split. The model operates on local $120\times120\times120$ voxel crops extracted from the native $200^3$ detector and receives a compact context token summarising crop geometry and detector-boundary position, both derived deterministically from the same voxel hits. This preserves containment information without relying on fitted trajectories or external reconstruction.

The second target is PILArNet~\cite{adams2020pilarnet}. Here the detector technology changes to a LArTPC, and the downstream tasks are five-class single-particle and multi-particle classification. We use the public $768^3$-pixel release throughout and construct a fixed 80k/2k/18k train/validation/test event split from the public HDF5 files. Single-particle classification uses centred particle crops with a compact voxel-derived metadata branch. Multi-particle classification adds a lightweight context transformer over the shared transferred encoder to aggregate particles from the same event, together with voxel-derived per-particle context features. These metadata branches are included to compensate for cropping or rescaling, not to inject external detector information. For fair comparison with the published literature, transfer performance is reported with micro-accuracy and the entropy-based AUROC definition used by Ref.~\cite{koh2023uncertainty}.

\subsection{\label{sec:evaluation_protocol}Evaluation protocol}

The main in-domain classification and regression comparisons use all three downstream variants defined at the start of the Results. Studies that currently compare only \scratchmodel{} and \maerelmodel{} are interpreted as tests of the full pre-training recipe relative to random initialisation, not as measurements of the relational contribution. Classification performance is reported with one-vs-rest operating curves and confusion matrices, together with threshold scans based on the figure of merit $\mathrm{FOM}=S/\sqrt{S+B}$, where $S$ and $B$ denote the expected signal and background yields after selection. Regression performance is summarised through selected-sample residual distributions and vertex-error distributions. Transfer benchmarks use fixed public splits, with the scintillator validation hold-out and the PILArNet manifests defined above.

The data-efficiency study keeps the validation and test sets fixed, varies only the number of labelled training events, and repeats each setting across three random seeds. This isolates the extent to which pre-training reduces labelled-data requirements. Exact manifests, checkpoint-selection rules and implementation details are provided in the released code, while technical details that would interrupt the main narrative are summarised in the Supplementary Information.

\begin{acknowledgments}
Neural network training was supported through the Swiss AI Initiative via a grant from the Swiss National Supercomputing Centre (CSCS), project ID a0178, on the Alps system.
\end{acknowledgments}

\section*{Data availability}

The simulated \fasercal{} dataset underlying the main results of this manuscript was produced with the publicly available \fasercal{} simulation and reconstruction framework, available at \url{https://github.com/rubbiaa/FASER}.

The datasets used for the transfer-learning studies are already public. The fine-grained plastic-scintillator benchmark is available on Zenodo at \url{https://doi.org/10.5281/zenodo.7347562}. The PILArNet data used for the LArTPC transfer study are available through the public PILArNet project at \url{https://osf.io/bu4fp/}, including the LArTPC simulation component at \url{https://osf.io/vruzp/}.

\section*{Code availability}

All code used in this study is available at \url{https://github.com/saulam/faserDL}.

\section*{Competing interests}

The authors declare no competing interests.

\bibliography{main}

@techreport{CERN-FASER-NOTE-2026-004,
  author       = {Alonso-Monsalve, S. and Cavanagh, C. and Cufino, F. and Kose, U. and
                  Masciellani, A. and Rubbia, A. and Sgalaberna, D. and Villa, E. and
                  Zhao, X. and Axiotis, K. and others},
  title        = {FASERCal Conceptual Design Report},
  institution  = {CERN},
  number       = {CERN-FASER-NOTE-2026-004},
  year         = {2026},
  url          = {https://cds.cern.ch/record/2954673},
}

@article{abreu2020detecting,
  author       = {{FASER Collaboration}},
  title        = {Detecting and studying high-energy collider neutrinos with {FASER} at the {LHC}},
  journal      = {The European Physical Journal C},
  volume       = {80},
  number       = {1},
  pages        = {61},
  year         = {2020},
  doi          = {10.1140/epjc/s10052-020-7631-5},
  url          = {https://doi.org/10.1140/epjc/s10052-020-7631-5}
}

@article{abreu2023firstdirect,
  author       = {{FASER Collaboration}},
  title        = {First Direct Observation of Collider Neutrinos with {FASER} at the {LHC}},
  journal      = {Physical Review Letters},
  volume       = {131},
  number       = {3},
  pages        = {031801},
  year         = {2023},
  doi          = {10.1103/PhysRevLett.131.031801},
  url          = {https://doi.org/10.1103/PhysRevLett.131.031801}
}

@article{abraham2024crosssections,
  author       = {{FASER Collaboration}},
  title        = {First Measurement of the {$\nu_e$} and {$\nu_\mu$} Interaction Cross Sections at the {LHC} with {FASER}'s Emulsion Detector},
  journal      = {Physical Review Letters},
  volume       = {133},
  number       = {2},
  pages        = {021802},
  year         = {2024},
  doi          = {10.1103/PhysRevLett.133.021802},
  url          = {https://doi.org/10.1103/PhysRevLett.133.021802}
}

@article{cruzmartinez2024lhcneutrinoion,
  author       = {Cruz-Martinez, Juan M. and Fieg, Max and Giani, Tommaso and Krack, Peter and M{\"a}kel{\"a}, Toni and Rabemananjara, Tanjona R. and Rojo, Juan},
  title        = {The {LHC} as a Neutrino-Ion Collider},
  journal      = {The European Physical Journal C},
  volume       = {84},
  pages        = {369},
  year         = {2024},
  doi          = {10.1140/epjc/s10052-024-12665-1},
  url          = {https://doi.org/10.1140/epjc/s10052-024-12665-1}
}

@article{aurisano2016cvn,
  title         = {A Convolutional Neural Network Neutrino Event Classifier},
  author        = {Aurisano, A. and Radovic, A. and Rocco, D. and Himmel, A. and Messier, M. D. and Niner, E. and Pawloski, G. and Psihas, F. and Sousa, A. and Vahle, P.},
  journal       = {JINST},
  volume        = {11},
  number        = {09},
  pages         = {P09001},
  year          = {2016},
  doi           = {10.1088/1748-0221/11/09/P09001},
  eprint        = {1604.01444},
  archivePrefix = {arXiv},
  primaryClass  = {hep-ex}
}

@article{radovic2018mlfrontiers,
  author  = {Radovic, Alexander and Williams, Mike and Rousseau, David and Kagan, Michael and Bonacorsi, Daniele and Himmel, Alexander and Aurisano, Adam and Terao, Kazuhiro and Wongjirad, Taritree},
  title   = {Machine learning at the energy and intensity frontiers of particle physics},
  journal = {Nature},
  volume  = {560},
  number  = {7716},
  pages   = {41--48},
  year    = {2018},
  doi     = {10.1038/s41586-018-0361-2},
  url     = {https://doi.org/10.1038/s41586-018-0361-2}
}

@article{karagiorgi2022mlnewphysics,
  author  = {Karagiorgi, Georgia and Kasieczka, Gregor and Kravitz, Scott and Nachman, Benjamin and Shih, David},
  title   = {Machine learning in the search for new fundamental physics},
  journal = {Nature Reviews Physics},
  volume  = {4},
  pages   = {399--412},
  year    = {2022},
  doi     = {10.1038/s42254-022-00441-9},
  url     = {https://doi.org/10.1038/s42254-022-00441-9}
}

@article{psihas2020review,
  author  = {Psihas, Fernanda and Groh, Micah and Tunnell, Christopher and Warburton, Karl},
  title   = {A review on machine learning for neutrino experiments},
  journal = {International Journal of Modern Physics A},
  volume  = {35},
  number  = {33},
  pages   = {2043005},
  year    = {2020},
  doi     = {10.1142/S0217751X20430058},
  url     = {https://doi.org/10.1142/S0217751X20430058}
}

@article{Babicz2022Adversarial,
  author  = {Babicz, M. and Alonso-Monsalve, S. and Dolan, S. and Terao, K.},
  title   = {Adversarial methods to reduce simulation bias in neutrino interaction event filtering at liquid argon time projection chambers},
  journal = {Phys. Rev. D},
  volume  = {105},
  pages   = {112009},
  year    = {2022},
  doi     = {10.1103/PhysRevD.105.112009},
  url     = {https://doi.org/10.1103/PhysRevD.105.112009}
}

@article{Harris2025RS3L,
  author    = {Harris, P. and Krupa, J. and Kagan, M. and Maier, B. and Woodward, N.},
  title     = {Resimulation-based self-supervised learning for pretraining physics foundation models},
  journal   = {Phys. Rev. D},
  volume    = {111},
  pages     = {032010},
  year      = {2025},
  doi       = {10.1103/PhysRevD.111.032010},
  url       = {https://doi.org/10.1103/PhysRevD.111.032010}
}

@article{Heinrich2024MPM,
  author    = {Heinrich, Lukas and Golling, Tobias and Kagan, Michael and Klein, Steven and Leigh, Matthew and Osadchy, Margarita and Raine, John A.},
  title     = {Masked particle modeling on sets: towards self-supervised high energy physics foundation models},
  journal   = {Machine Learning: Science and Technology},
  volume    = {5},
  number    = {3},
  pages     = {035074},
  year      = {2024},
  doi       = {10.1088/2632-2153/ad64a8},
  url       = {https://doi.org/10.1088/2632-2153/ad64a8}
}

@article{Birk2024OmniJet,
  author    = {Birk, Joschka and Hallin, Anna and Kasieczka, Gregor},
  title     = {OmniJet-$\\alpha$: the first cross-task foundation model for particle physics},
  journal   = {Machine Learning: Science and Technology},
  volume    = {5},
  number    = {3},
  pages     = {035031},
  year      = {2024},
  doi       = {10.1088/2632-2153/ad66ad},
  url       = {https://doi.org/10.1088/2632-2153/ad66ad}
}

@article{Vigl2024Finetuning,
  author    = {Vigl, Matthias and Hartman, Nicole and Heinrich, Lukas},
  title     = {Finetuning foundation models for joint analysis optimization},
  journal   = {Machine Learning: Science and Technology},
  volume    = {5},
  number    = {2},
  pages     = {025075},
  year      = {2024},
  doi       = {10.1088/2632-2153/ad55a3},
  url       = {https://doi.org/10.1088/2632-2153/ad55a3}
}

@article{acciarri2017microboonecNN,
  title         = {Convolutional Neural Networks Applied to Neutrino Events in a Liquid Argon Time Projection Chamber},
  author        = {Acciarri, R. and others},
  collaboration = {MicroBooNE Collaboration},
  journal       = {JINST},
  volume        = {12},
  number        = {03},
  pages         = {P03011},
  year          = {2017},
  doi           = {10.1088/1748-0221/12/03/P03011},
  eprint        = {1611.05531},
  archivePrefix = {arXiv},
  primaryClass  = {physics.ins-det}
}

@article{abi2020dunecvn,
  title         = {Neutrino interaction classification with a convolutional neural network in the {DUNE} far detector},
  author        = {Abi, B. and others},
  collaboration = {DUNE Collaboration},
  journal       = {Phys. Rev. D},
  volume        = {102},
  pages         = {092003},
  year          = {2020},
  doi           = {10.1103/PhysRevD.102.092003},
  eprint        = {2006.15052},
  archivePrefix = {arXiv},
  primaryClass  = {physics.ins-det}
}

@article{domine2020sparse,
  title         = {Scalable deep convolutional neural networks for sparse, locally dense liquid argon time projection chamber data},
  author        = {Domin{\'e}, Laura and Terao, Kazuhiro},
  journal       = {Phys. Rev. D},
  volume        = {102},
  number        = {1},
  pages         = {012005},
  year          = {2020},
  doi           = {10.1103/PhysRevD.102.012005},
  eprint        = {1903.05663},
  archivePrefix = {arXiv},
  primaryClass  = {hep-ex}
}

@article{PhysRevD.103.032005,
  title = {Graph neural network for {3D} classification of ambiguities and optical crosstalk in scintillator-based neutrino detectors},
  author = {Alonso-Monsalve, Sa\'ul and Douqa, Dana and Jes\'us-Valls, C\'esar and Lux, Thorsten and Pina-Otey, Sebastian and S\'anchez, Federico and Sgalaberna, Davide and Whitehead, Leigh H.},
  journal = {Phys. Rev. D},
  volume = {103},
  issue = {3},
  pages = {032005},
  numpages = {24},
  year = {2021},
  month = {Feb},
  publisher = {American Physical Society},
  doi = {10.1103/PhysRevD.103.032005},
  url = {https://link.aps.org/doi/10.1103/PhysRevD.103.032005}
}

@article{abratenko2021sparsessnet,
  title         = {Semantic Segmentation with a Sparse Convolutional Neural Network for Event Reconstruction in {MicroBooNE}},
  author        = {Abratenko, P. and others},
  collaboration = {MicroBooNE Collaboration},
  journal       = {Phys. Rev. D},
  volume        = {103},
  pages         = {052012},
  year          = {2021},
  doi           = {10.1103/PhysRevD.103.052012},
  eprint        = {2012.08513},
  archivePrefix = {arXiv},
  primaryClass  = {hep-ex}
}

@article{alonsoMonsalve2024overlap,
  title   = {Deep-learning-based decomposition of overlapping-sparse images: application at the vertex of simulated neutrino interactions},
  author  = {Alonso-Monsalve, Sa{\'u}l and Sgalaberna, Davide and Zhao, Xingyu and Molines, Adrien and McGrew, Clark and Rubbia, Andr{\'e}},
  journal = {Communications Physics},
  volume  = {7},
  pages   = {173},
  year    = {2024},
  doi     = {10.1038/s42005-024-01669-8}
}

@article{coelho2021quantization,
  author  = {Coelho, Claudionor N. and Kuusela, Aki and Li, Shan and Zhuang, Hao and Ngadiuba, Jennifer and Aarrestad, Thea Klaeboe and Loncar, Vladimir and Pierini, Maurizio and Pol, Adrian Alan and Summers, Sioni},
  title   = {Automatic heterogeneous quantization of deep neural networks for low-latency inference on the edge for particle detectors},
  journal = {Nature Machine Intelligence},
  volume  = {3},
  pages   = {675--686},
  year    = {2021},
  doi     = {10.1038/s42256-021-00356-5},
  url     = {https://doi.org/10.1038/s42256-021-00356-5}
}

@article{govorkova2022autoencoders,
  author  = {Govorkova, Ekaterina and Puljak, Ema and Aarrestad, Thea Klaeboe and James, Thomas and Loncar, Vladimir and Pierini, Maurizio and Pol, Adrian Alan and Summers, Sioni and Ngadiuba, Jennifer and Nguyen, Thong Q. and Duarte, Javier and Wu, Zhenbin and others},
  title   = {Autoencoders on field-programmable gate arrays for real-time, unsupervised new physics detection at 40 {MHz} at the {Large Hadron Collider}},
  journal = {Nature Machine Intelligence},
  volume  = {4},
  pages   = {154--161},
  year    = {2022},
  doi     = {10.1038/s42256-022-00441-3},
  url     = {https://doi.org/10.1038/s42256-022-00441-3}
}

@article{pata2024particleflow,
  author  = {Pata, Joosep and Wulff, Eric and Mokhtar, Farouk and Southwick, David and Zhang, Mengke and Girone, Maria and Duarte, Javier},
  title   = {Improved particle-flow event reconstruction with scalable neural networks for current and future particle detectors},
  journal = {Communications Physics},
  volume  = {7},
  pages   = {107},
  year    = {2024},
  doi     = {10.1038/s42005-024-01599-5},
  url     = {https://doi.org/10.1038/s42005-024-01599-5}
}

@article{belis2024quantum,
  author  = {Belis, Vasilis and Wo{\'z}niak, Kinga Anna and Puljak, Ema and others},
  title   = {Quantum anomaly detection in the latent space of proton collision events at the {LHC}},
  journal = {Communications Physics},
  volume  = {7},
  pages   = {334},
  year    = {2024},
  doi     = {10.1038/s42005-024-01811-6},
  url     = {https://doi.org/10.1038/s42005-024-01811-6}
}

@inproceedings{graham2018submanifold,
  author       = {Graham, Benjamin and Engelcke, Martin and van der Maaten, Laurens},
  title        = {{3D} Semantic Segmentation with Submanifold Sparse Convolutional Networks},
  booktitle    = {Proceedings of the IEEE/CVF Conference on Computer Vision and Pattern Recognition},
  pages        = {9224--9232},
  year         = {2018},
  doi          = {10.1109/CVPR.2018.00962},
  url          = {https://openaccess.thecvf.com/content_cvpr_2018/html/Graham_3D_Semantic_Segmentation_CVPR_2018_paper.html}
}

@misc{spconv2020,
  author = {Yan, Yan},
  title  = {Spconv: Spatially Sparse Convolution Library},
  year   = {2020},
  publisher = {GitHub},
  journal = {GitHub repository},
  howpublished = {\url{https://github.com/traveller59/spconv}}
}

@inproceedings{devlin-etal-2019-bert,
    title = "{BERT}: Pre-training of Deep Bidirectional Transformers for Language Understanding",
    author = "Devlin, Jacob  and
      Chang, Ming-Wei  and
      Lee, Kenton  and
      Toutanova, Kristina",
    editor = "Burstein, Jill  and
      Doran, Christy  and
      Solorio, Thamar",
    booktitle = "Proceedings of the 2019 Conference of the North {A}merican Chapter of the Association for Computational Linguistics: Human Language Technologies, Volume 1 (Long and Short Papers)",
    month = jun,
    year = "2019",
    address = "Minneapolis, Minnesota",
    publisher = "Association for Computational Linguistics",
    url = "https://aclanthology.org/N19-1423/",
    doi = "10.18653/v1/N19-1423",
    pages = "4171--4186"
}

@inproceedings{dosovitskiy2021vit,
    title={An Image is Worth 16x16 Words: Transformers for Image Recognition at Scale},
    author={Alexey Dosovitskiy and Lucas Beyer and Alexander Kolesnikov and Dirk Weissenborn and Xiaohua Zhai and Thomas Unterthiner and Mostafa Dehghani and Matthias Minderer and Georg Heigold and Sylvain Gelly and Jakob Uszkoreit and Neil Houlsby},
    booktitle={International Conference on Learning Representations},
    year={2021},
    url={https://openreview.net/forum?id=YicbFdNTTy}
}

@INPROCEEDINGS{he2022masked,
    author = {He, Kaiming and Chen, Xinlei and Xie, Saining and Li, Yanghao and Doll\'ar, Piotr and Girshick, Ross},
    title = {Masked Autoencoders Are Scalable Vision Learners},
    booktitle = {Proceedings of the IEEE/CVF Conference on Computer Vision and Pattern Recognition (CVPR)},
    month = {June},
    year = {2022},
    pages = {16000-16009}
}

@InProceedings{Xie_2022_CVPR,
    author = {Xie, Zhenda and Zhang, Zheng and Cao, Yue and Lin, Yutong and Bao, Jianmin and Yao, Zhuliang and Dai, Qi and Hu, Han},
    title = {{SimMIM}: A Simple Framework for Masked Image Modeling},
    booktitle = {Proceedings of the IEEE/CVF Conference on Computer Vision and Pattern Recognition (CVPR)},
    month = {June},
    year = {2022},
    pages = {9653-9663}
}

@InProceedings{pmlr-v162-baevski22a,
    title = {data2vec: A General Framework for Self-supervised Learning in Speech, Vision and Language},
    author = {Baevski, Alexei and Hsu, Wei-Ning and Xu, Qiantong and Babu, Arun and Gu, Jiatao and Auli, Michael},
    booktitle = {Proceedings of the 39th International Conference on Machine Learning},
    pages = {1298--1312},
    year = {2022},
    editor = {Chaudhuri, Kamalika and Jegelka, Stefanie and Song, Le and Szepesvari, Csaba and Niu, Gang and Sabato, Sivan},
    volume = {162},
    series = {Proceedings of Machine Learning Research},
    month = {17--23 Jul},
    publisher = {PMLR},
    url = {https://proceedings.mlr.press/v162/baevski22a.html}
}

@InProceedings{Yu_2022_CVPR,
    author = {Yu, Xumin and Tang, Lulu and Rao, Yongming and Huang, Tiejun and Zhou, Jie and Lu, Jiwen},
    title = {{Point-BERT}: Pre-Training 3D Point Cloud Transformers With Masked Point Modeling},
    booktitle = {Proceedings of the IEEE/CVF Conference on Computer Vision and Pattern Recognition (CVPR)},
    month = {June},
    year = {2022},
    pages = {19313-19322}
}

@article{hou2022graphmae,
    title = {{GraphMAE}: Self-Supervised Masked Graph Autoencoders},
    author = {Hou, Zhenyu and Liu, Xiao and Cen, Yukuo and Dong, Yuxiao and Yang, Hongxia and Wang, Chunjie and Tang, Jie},
    journal = {arXiv preprint arXiv:2205.10803},
    year = {2022},
    doi = {10.48550/arXiv.2205.10803},
    url = {https://arxiv.org/abs/2205.10803}
}

@inproceedings{jaegle2021perceiver,
      title={{Perceiver IO}: A General Architecture for Structured Inputs \& Outputs}, 
      author={Andrew Jaegle and Sebastian Borgeaud and Jean-Baptiste Alayrac and Carl Doersch and Catalin Ionescu and David Ding and Skanda Koppula and Daniel Zoran and Andrew Brock and Evan Shelhamer and Olivier Hénaff and Matthew M. Botvinick and Andrew Zisserman and Oriol Vinyals and Joāo Carreira},
      year={2022},
      eprint={2107.14795},
      archivePrefix={arXiv},
      primaryClass={cs.LG},
      url={https://arxiv.org/abs/2107.14795}, 
}

@inproceedings{kendall2018multi,
  author={Cipolla, Roberto and Gal, Yarin and Kendall, Alex},
  booktitle={2018 IEEE/CVF Conference on Computer Vision and Pattern Recognition}, 
  title={Multi-task Learning Using Uncertainty to Weigh Losses for Scene Geometry and Semantics}, 
  year={2018},
  volume={},
  number={},
  pages={7482-7491},
  doi={10.1109/CVPR.2018.00781}
}

@inproceedings{loshchilov2019adamw,
    title={Decoupled Weight Decay Regularization},
    author={Ilya Loshchilov and Frank Hutter},
    booktitle={International Conference on Learning Representations},
    year={2019},
    url={https://openreview.net/forum?id=Bkg6RiCqY7},
}

@inproceedings{bao2022beit,
    title={{BE}iT: {BERT} Pre-Training of Image Transformers},
    author={Hangbo Bao and Li Dong and Songhao Piao and Furu Wei},
    booktitle={International Conference on Learning Representations},
    year={2022},
    url={https://openreview.net/forum?id=p-BhZSz59o4}
}

@article{polyak1992ema,
    author = {Polyak, B. T. and Juditsky, A. B.},
    title = {Acceleration of Stochastic Approximation by Averaging},
    journal = {SIAM Journal on Control and Optimization},
    volume = {30},
    number = {4},
    pages = {838-855},
    year = {1992},
    doi = {10.1137/0330046},
    URL = { 
            https://doi.org/10.1137/0330046
    },
    eprint = {     
            https://doi.org/10.1137/0330046
    }
    ,
}

@misc{adams2020pilarnet,
  title         = {{PILArNet}: Public Dataset for Particle Imaging Liquid Argon Detectors in High Energy Physics},
  author        = {Adams, C. and Terao, K. and Wongjirad, T.},
  year          = {2020},
  eprint        = {2006.01993},
  archivePrefix = {arXiv},
  primaryClass  = {physics.ins-det},
  url           = {https://arxiv.org/abs/2006.01993}
}

@article{drielsma2020grappa,
  title = {Clustering of electromagnetic showers and particle interactions with graph neural networks in liquid argon time projection chambers},
  author = {Drielsma, Fran\ifmmode \mbox{\c{c}}\else \c{c}\fi{}ois and Lin, Qing and de Soux, Pierre C\^ote and Domin\'e, Laura and Itay, Ran and Koh, Dae Heun and Nelson, Bradley J. and Terao, Kazuhiro and Tsang, Ka Vang and Usher, Tracy L.},
  collaboration = {DeepLearnPhysics Collaboration},
  journal = {Phys. Rev. D},
  volume = {104},
  issue = {7},
  pages = {072004},
  numpages = {19},
  year = {2021},
  month = {Oct},
  publisher = {American Physical Society},
  doi = {10.1103/PhysRevD.104.072004},
  url = {https://link.aps.org/doi/10.1103/PhysRevD.104.072004}
}

@article{AlonsoMonsalve2023TrackFitting,
  author       = {Alonso-Monsalve, Sa{\'u}l and Sgalaberna, Davide and Zhao, Xingyu and McGrew, Clark and Rubbia, Andr{\'e}},
  title        = {Artificial intelligence for improved fitting of trajectories of elementary particles in dense materials immersed in a magnetic field},
  journal      = {Communications Physics},
  volume       = {6},
  pages        = {119},
  year         = {2023},
  doi          = {10.1038/s42005-023-01239-4},
  url          = {https://doi.org/10.1038/s42005-023-01239-4}
}

@article{koh2023uncertainty,
  author       = {Koh, D. H. and Mishra, A. and Terao, K.},
  title        = {Deep neural network uncertainty quantification for {LArTPC} reconstruction},
  journal      = {Journal of Instrumentation},
  volume       = {18},
  number       = {12},
  pages        = {P12013},
  year         = {2023},
  doi          = {10.1088/1748-0221/18/12/P12013},
  url          = {https://doi.org/10.1088/1748-0221/18/12/P12013}
}

@article{Chappell2022Application,
  author  = {Andrew Chappell and Leigh H. Whitehead},
  title   = {Application of transfer learning to neutrino interaction classification},
  journal = {The European Physical Journal C},
  volume  = {82},
  pages   = {1099},
  year    = {2022},
  doi     = {10.1140/epjc/s10052-022-11066-6},
  url     = {https://doi.org/10.1140/epjc/s10052-022-11066-6}
}

@article{Bonilla2026Transfer,
  author    = {Jos{\'e} L. Bonilla and Krzysztof M. Graczyk and Artur M. Ankowski and Rwik Dharmapal Banerjee and Beata E. Kowal and Hemant Prasad and Jan T. Sobczyk},
  title     = {Transfer learning for neutrino scattering: Domain adaptation with generative adversarial networks},
  journal   = {Physical Review D},
  volume    = {113},
  number    = {5},
  pages     = {053001},
  year      = {2026},
  month     = mar,
  publisher = {American Physical Society},
  doi       = {10.1103/kwjj-wp1c},
  url       = {https://link.aps.org/doi/10.1103/kwjj-wp1c}
}

@article{Wilkinson2025Contrastive,
  author    = {Alex Wilkinson and Radi Radev and Sa{\'u}l Alonso-Monsalve},
  title     = {Contrastive learning for robust representations of neutrino data},
  journal   = {Physical Review D},
  volume    = {111},
  number    = {9},
  pages     = {092011},
  year      = {2025},
  month     = may,
  publisher = {American Physical Society},
  doi       = {10.1103/PhysRevD.111.092011},
  url       = {https://link.aps.org/doi/10.1103/PhysRevD.111.092011}
}

@article{Young2026Particle,
  author    = {Samuel Young and Yeon-jae Jwa and Kazuhiro Terao},
  title     = {Particle trajectory representation learning with masked point modeling},
  journal   = {Machine Learning: Science and Technology},
  year      = {2026},
  month     = feb,
  doi       = {10.1088/2632-2153/ae47b8},
  url       = {https://doi.org/10.1088/2632-2153/ae47b8}
}

@misc{babicz2026transformerbasedpulseshapediscrimination,
      title={Transformer-Based Pulse Shape Discrimination in {HPGe} Detectors with Masked Autoencoder Pre-training}, 
      author={Marta Babicz and Saúl Alonso-Monsalve and Alain Fauquex and Laura Baudis},
      year={2026},
      eprint={2603.06192},
      archivePrefix={arXiv},
      primaryClass={hep-ex},
      url={https://arxiv.org/abs/2603.06192}, 
}

@inproceedings{Yu:2025t0,
  author    = {Yu, Felix and Kamp, Nicholas and Arg{\"u}elles, Carlos},
  title     = {Reducing Simulation Dependence in Neutrino Telescopes with Masked Point Modeling},
  doi       = {10.22323/1.501.1218},
  booktitle = {Proceedings of 39th International Cosmic Ray Conference --- PoS(ICRC2025)},
  year      = {2025},
  volume    = {501},
  pages     = {1218}
}

@misc{Sagar2025Adapting,
  title         = {Adapting Vision-Language Models for Neutrino Event Classification in High-Energy Physics},
  author        = {Dikshant Sagar and Kaiwen Yu and Alejandro Yankelevich and Jianming Bian and Pierre Baldi},
  year          = {2025},
  eprint        = {2509.08461},
  archivePrefix = {arXiv},
  primaryClass  = {cs.LG},
  url           = {https://arxiv.org/abs/2509.08461}
}

@misc{Albert2025Deep,
  title         = {Deep Learning Framework for Enhanced Neutrino Reconstruction of Single-line Events in the {ANTARES} Telescope},
  author        = {A. Albert and S. Alves and M. Andr{\'e} and others},
  year          = {2025},
  eprint        = {2511.16614},
  archivePrefix = {arXiv},
  primaryClass  = {physics.comp-ph},
  url           = {https://arxiv.org/abs/2511.16614}
}

@article{andreopoulos2010genie,
  author  = {Andreopoulos, C. and Bell, A. and Bhattacharya, D. and Cavanna, F. and
             Dytman, S. and Gallagher, H. and Guzowski, P. and Hatcher, R. and
             Kehayias, P. and Meregaglia, A. and Naples, D. and Pearce, G. and
             Poskanzer, A. and Raboanary, R. and Technical, A. and Wilking, M. and others},
  title   = {The {GENIE} neutrino {Monte Carlo} generator},
  journal = {Nuclear Instruments and Methods in Physics Research Section A},
  volume  = {614},
  number  = {1},
  pages   = {87--104},
  year    = {2010},
  doi     = {10.1016/j.nima.2009.12.009},
}

@article{bierlich2022pythia8,
  author  = {Bierlich, Christian and Chakraborty, Smita and Desai, Nishita and Gellersen, Leif and Helenius, Ilkka and Ilten, Philip and L{\"o}nnblad, Leif and Mrenna, Stephen and Prestel, Stefan and Preuss, Christian Thomas and Sj{\"o}strand, Torbj{\"o}rn and Skands, Peter and Utheim, Marius and Verber{\'e}k, Rob},
  title   = {A comprehensive guide to the physics and usage of {PYTHIA~8.3}},
  journal = {SciPost Physics Codebases},
  volume  = {8},
  pages   = {r8.3},
  year    = {2022},
  doi     = {10.21468/SciPostPhysCodeb.8.3},
}

@article{agostinelli2003geant4,
  author  = {Agostinelli, S. and Allison, J. and Amako, K. and Apostolakis, J. and Araujo, H. and Arce, P. and Asai, M. and Axen, D. and Banerjee, S. and Barrand, G. and others},
  title   = {{Geant4}: a simulation toolkit},
  journal = {Nuclear Instruments and Methods in Physics Research Section A},
  volume  = {506},
  number  = {3},
  pages   = {250--303},
  year    = {2003},
  doi     = {10.1016/S0168-9002(03)01368-8},
}

@article{FASER:2024ref,
    author = "Mammen Abraham, Roshan and others",
    collaboration = "FASER",
    title = "{First Measurement of the Muon Neutrino Interaction Cross Section and Flux as a Function of Energy at the LHC with FASER}",
    eprint = "2412.03186",
    archivePrefix = "arXiv",
    primaryClass = "hep-ex",
    reportNumber = "CERN-EP-2024-309",
    doi = "10.1103/PhysRevLett.134.211801",
    journal = "Phys. Rev. Lett.",
    volume = "134",
    number = "21",
    pages = "211801",
    year = "2025"
}

@article{SNDLHC2023,
  author        = {{SND@LHC Collaboration}},
  title         = {Observation of collider muon neutrinos with the {SND@LHC} experiment},
  journal       = {Physical Review Letters},
  volume        = {131},
  number        = {3},
  pages         = {031802},
  year          = {2023},
  doi           = {10.1103/PhysRevLett.131.031802},
  eprint        = {2305.09383},
  archiveprefix = {arXiv},
  primaryclass  = {hep-ex},
  note          = {CERN-EP-2023-092},
  url           = {https://arxiv.org/abs/2305.09383},
}

@inproceedings{fan2017pointset,
    author = {Fan, Haoqiang and Su, Hao and Guibas, Leonidas J.},
    title = {A Point Set Generation Network for 3D Object Reconstruction from a Single Image},
    booktitle = {Proceedings of the IEEE Conference on Computer Vision and Pattern Recognition},
    pages = {605--613},
    year = {2017},
    doi = {10.1109/CVPR.2017.72},
    url = {https://doi.org/10.1109/CVPR.2017.72}
}

@article{golan2012nuwro,
  author  = {Golan, Tomasz and Juszczak, Cezary and Sobczyk, Jan T.},
  title   = {Final State Interactions Effects in Neutrino-Nucleus Interactions},
  journal = {Physical Review C},
  volume  = {86},
  pages   = {015505},
  year    = {2012},
  doi     = {10.1103/PhysRevC.86.015505},
}

@article{sjostrand2006pythia6,
  author        = {Sj{\"o}strand, Torbj{\"o}rn and Mrenna, Stephen and Skands, Peter Z.},
  title         = {{PYTHIA 6.4 Physics and Manual}},
  journal       = {Journal of High Energy Physics},
  volume        = {2006},
  number        = {05},
  pages         = {026},
  year          = {2006},
  doi           = {10.1088/1126-6708/2006/05/026},
  eprint        = {hep-ph/0603175},
  archivePrefix = {arXiv},
  primaryClass  = {hep-ph}
}

\appendix
\renewcommand{\thefigure}{\arabic{figure}}
\renewcommand{\thetable}{\arabic{table}}
\setcounter{figure}{0}
\setcounter{table}{0}

\section{\label{app:detailed_clsreg}Supplementary classification and regression diagnostics}

Additional operating-point scans and class-resolved regression boxplots are shown here to provide the detailed diagnostics underlying the compact classification and regression summary in Fig.~\ref{fig:classification_regression_results}.

\begin{figure*}[htbp]
  \centering
  \includegraphics[width=0.98\linewidth]{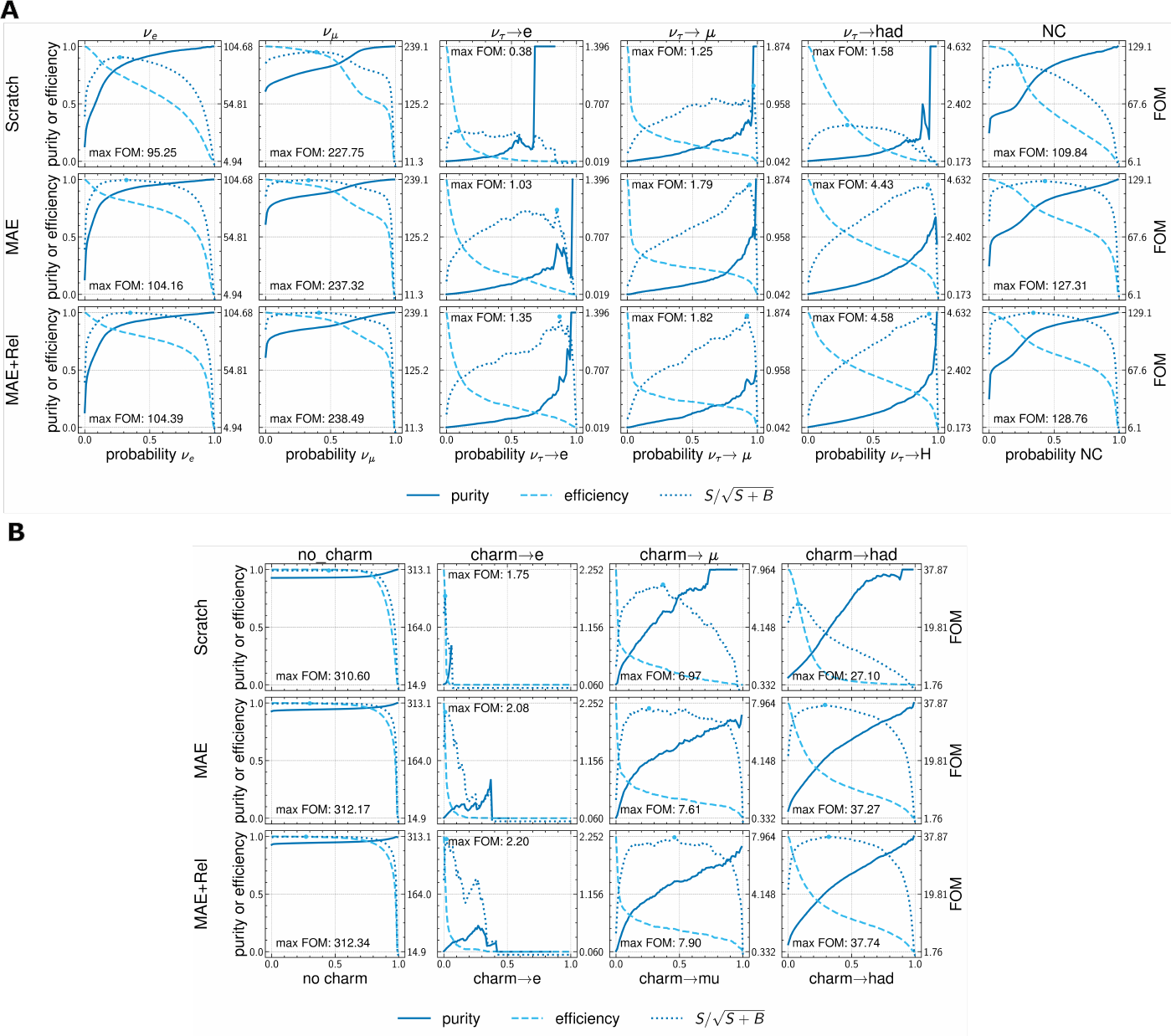}
  \caption{\textbf{Supplementary classification operating-point scans.}
  \textbf{A} Purity (solid), efficiency (dashed) and figure of merit $S/\sqrt{S+B}$ (dotted) as functions of the classifier score threshold for the six flavour categories. \textbf{B} The corresponding scans for the four charm categories. Rows compare \scratchmodel{}, \maemodel{} and \maerelmodel{}. These scans define the per-class thresholds used for the confusion matrices in Fig.~\ref{fig:classification_regression_results} and provide the maximum-FOM values quoted in the Results.}
  \label{fig:classification_supplement}
\end{figure*}

\begin{figure*}[htbp]
  \centering
  \includegraphics[width=0.98\linewidth]{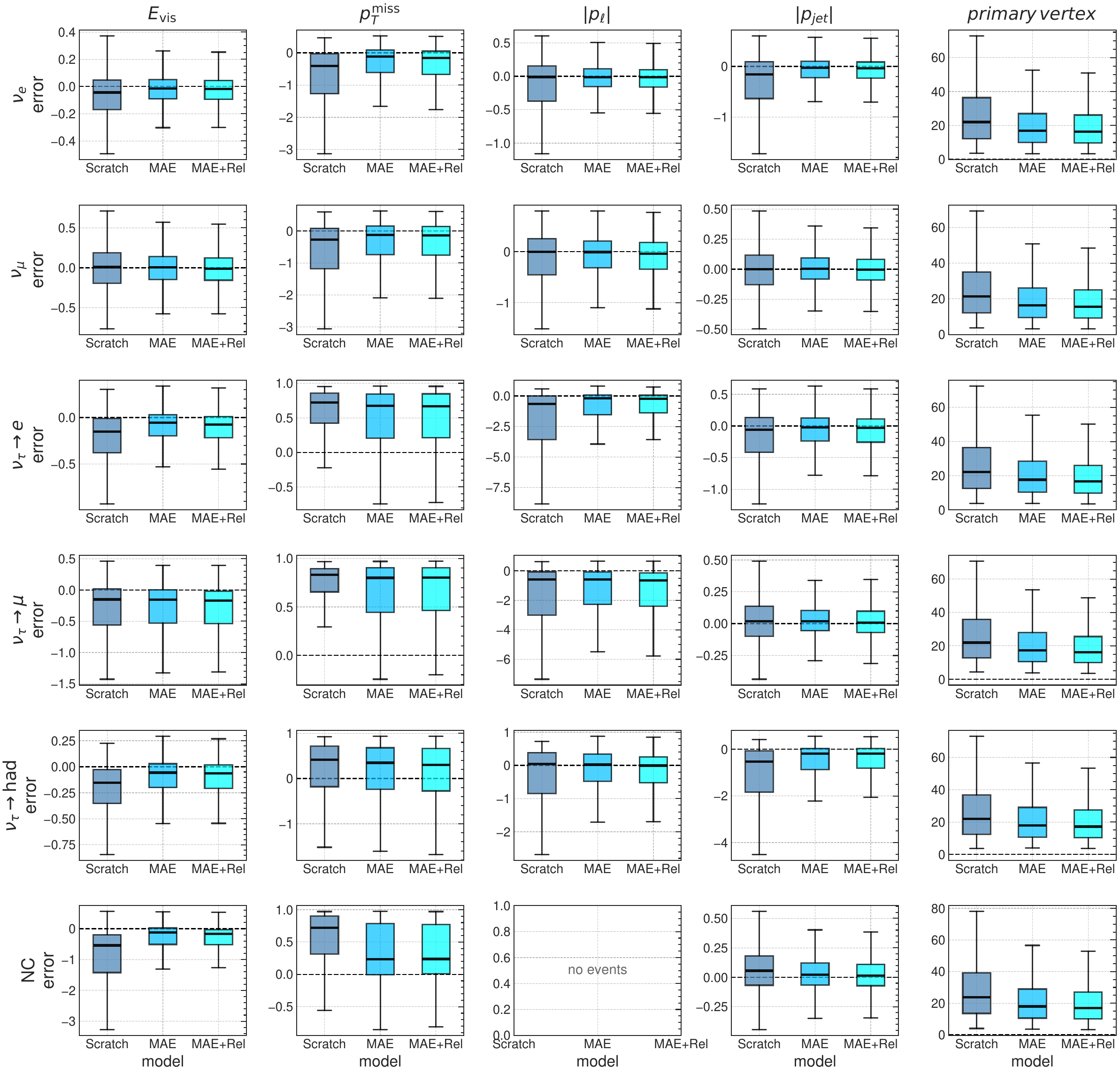}
  \caption{\textbf{Supplementary class-resolved regression boxplots.}
  Regression-error boxplots for the selected event sample, resolved by true flavour category and target observable. Columns show the fractional residuals in $E_{\mathrm{vis}}$, $p_T^{\mathrm{miss}}$, $|p_\ell|$ and $|p_{\mathrm{jet}}|$, defined as $(x^{\mathrm{true}}-x^{\mathrm{reco}})/x^{\mathrm{true}}$, together with the primary-vertex reconstruction error $d_{\mathrm{PV}}=\|\vec{x}_{\mathrm{PV}}^{\mathrm{true}}-\vec{x}_{\mathrm{PV}}^{\mathrm{reco}}\|$ in mm. Within each panel, the three boxplots compare \scratchmodel{}, \maemodel{} and \maerelmodel{} using the same convention as Fig.~\ref{fig:classification_regression_results}: boxes indicate the interquartile range, horizontal bars the median, and whiskers the central 95\% of the selected events. These class-resolved boxplots complement the all-class summary in Fig.~\ref{fig:classification_regression_results}.}
  \label{fig:regression_supplement}
\end{figure*}

\section{\label{app:architecture}Supplementary architectural details}

\begin{figure*}[htpb]
    \centering
    \includegraphics[width=1.0\textwidth]{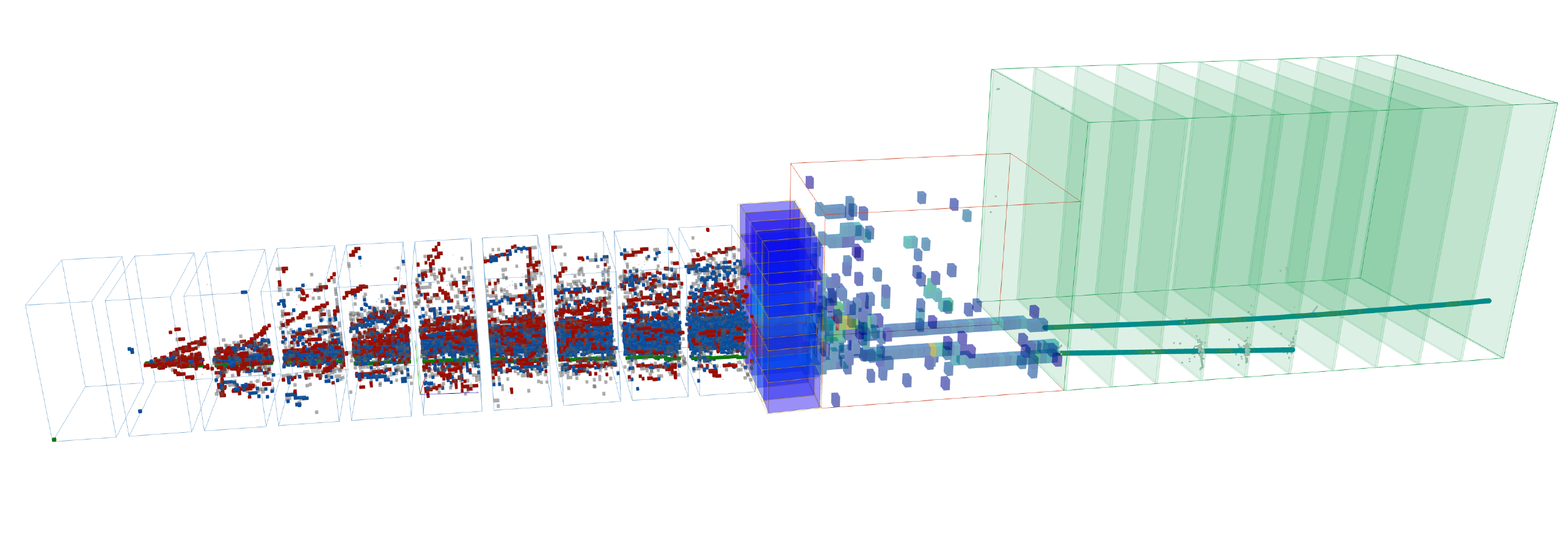}
    \caption{
    Event display of a simulated charged-current muon antineutrino interaction in the \fasercal{} detector.
    The event is a high-energy $\bar{\nu}_{\mu}$ CC interaction with a true visible energy of
    $E_{\mathrm{vis}} \simeq 1265~\mathrm{GeV}$ and a reconstructed primary-lepton momentum of
    $p_{\mu} \simeq 745~\mathrm{GeV}$.
    The neutrino interaction occurs inside the \threedcal{} and produces a dense shower with both
    hadronic and electromagnetic components, shown respectively by the reconstructed HAD voxels in red
    and EM voxels in blue. Voxels associated with the outgoing charged-current muon are shown in green.
    The display also shows the ghost voxels in grey.
    A long penetrating track, compatible with the outgoing muon, traverses the downstream \ecal{} and
    \ahcal{} before entering the muon spectrometer. A secondary muon is also produced and traverses the \ahcal{} and the muon spectrometer.
    }
    \label{fig:event_display_antinu_mu_cc}
\end{figure*}

\subsection{\label{app:tokenisation}Tokenisation and latent structure}

In the configuration used for the main experiments, the \threedcal{} is tokenised into $12 \times 12 \times 10$ voxel patches, yielding a $4 \times 4 \times 20$ patch grid. The detector is organised into ten modules, so each module spans two patch planes in depth. The \ahcal{} is tokenised into $6 \times 6 \times 5$ voxel patches, yielding a $3 \times 3 \times 8$ patch grid. Only occupied patches are propagated as active tokens, and occupancy masks are maintained through the attention blocks.

The \ecal{} is encoded as a compact token derived from its energy matrix. The muon spectrometer is encoded from fitted track summaries and an auxiliary presence or count signal. Learned detector-type embeddings distinguish the sources before Perceiver fusion.

\section{\label{app:training}Supplementary training details}

\subsection{\label{app:hyperparams}Hyperparameters}

Table~\ref{tab:hyperparams} summarises the key training hyperparameters for all model variants. The base learning rate is linearly scaled by the effective batch size divided by 256 inside the training framework; the values listed are the unscaled base rates. Per-step warm-up and cosine-annealing schedules are computed from the trainer state (number of devices, nodes and gradient-accumulation steps).

\begin{table*}[htbp]
\centering
\small
\setlength{\tabcolsep}{5pt}
\renewcommand{\arraystretch}{1.15}
\caption{\textbf{Training hyperparameters.} Pre-training proceeds in two phases: Phase~1 trains with masked reconstruction only (MAE), and Phase~2 continues from the Phase~1 checkpoint with an additional relational pass (MAE+Rel). Fine-tuning (FT) and Scratch use the same architecture and recipe except where noted.}
\label{tab:hyperparams}
\begin{tabular}{l c c c c}
\toprule
\textbf{Parameter} & \textbf{Phase 1 (MAE)} & \textbf{Phase 2 (MAE+Rel)} & \textbf{FT (pre-trained)} & \textbf{Scratch} \\
\midrule
Epochs                       & 400    & +100   & 20    & 40   \\
Batch size (per GPU)         & 512    & 512    & 1,024 & 1,024 \\
Base learning rate           & $10^{-4}$  & $10^{-4}$  & $5{\times}10^{-4}$ & $10^{-3}$ \\
Warm-up epochs               & 40     & 10     & 5     & 5 \\
Cosine-annealing epochs      & 360    & 90     & 15    & 35\\
Weight decay                 & 0.05   & 0.05   & 0.05  & 0.05 \\
$\beta_1 / \beta_2$          & 0.9 / 0.95 & 0.9 / 0.95 & 0.9 / 0.999 & 0.9 / 0.999 \\
Label smoothing              & 0.02   & 0.02   & 0.02  & 0.02 \\
Hit charge preprocessing   & log    & log    & log   & log  \\
\midrule
\multicolumn{5}{l}{\textit{Pre-training--specific}} \\
Mask ratio                   & 0.75   & 0.75   & ---   & ---  \\
Relational pass probability  & 0.0    & 0.5    & ---   & ---  \\
Relational mask ratio        & ---    & 0.25   & ---   & ---  \\
Reconstruction loss mode     & hybrid & hybrid & ---   & ---  \\
\midrule
\multicolumn{5}{l}{\textit{Fine-tuning--specific}} \\
Layer-wise learning-rate decay          & ---    & ---    & 0.75  & 1.0  \\
Exponential moving average decay                    & ---    & ---    & 0.9999 & 0.9999 \\
Drop-path rate               & 0.0    & 0.0    & 0.2   & 0.2  \\
Head init scale              & ---    & ---    & $2{\times}10^{-5}$ & $2{\times}10^{-5}$ \\
\midrule
\multicolumn{5}{l}{\textit{Hardware}} \\
GPUs                         & \multicolumn{2}{c}{8$\times$GH200 (2 nodes)} & \multicolumn{2}{c}{1$\times$H100} \\
\bottomrule
\end{tabular}
\end{table*}

\subsection{\label{app:model_scale}Model size and computational cost}

Table~\ref{tab:model_compute} reports the parameter counts and active compute cost of the final runs. The downstream count refers to the model used for fine-tuning and inference in the headline \fasercal{} experiments. The pre-training count includes the MAE decoder and relational heads, which are discarded before downstream fine-tuning.

\begin{table*}[htbp]
\centering
\footnotesize
\setlength{\tabcolsep}{4pt}
\renewcommand{\arraystretch}{1.15}
\caption{\textbf{Model size and active computational cost.} The downstream count refers to the model used for fine-tuning and inference in the headline \fasercal{} experiments. The pre-training count includes the masked-autoencoder decoder and relational heads, which are discarded before downstream fine-tuning.}
\label{tab:model_compute}

\begin{tabular}{l r r c r r}
\toprule
Stage & Trainable model parameters & Optimisation steps & Hardware & Active wall time & GPU-hours \\
\midrule
MAE pre-training & 34,258,301 & 101,600 & $8\times$GH200 & 73.0\,h & 584 \\
MAE+Rel continuation & 34,258,301 & 25,400 & $8\times$GH200 & 53.4\,h & 427 \\
\midrule
Total pre-training & 34,258,301 & 127,000 & $8\times$GH200 & 126.4\,h & 1,011 \\
Joint fine-tuning & 31,936,341 & 20,340 & $1\times$H100 & 4.1\,h & 4.1 \\
\bottomrule
\end{tabular}

\end{table*}

\subsection{\label{app:optimisation}Optimisation}

All stages use the AdamW optimiser~\cite{loshchilov2019adamw} with the parameters listed in Table~\ref{tab:hyperparams}. Fine-tuning further uses layer-wise learning-rate decay~\cite{bao2022beit}, exponential moving averages of the model parameters~\cite{polyak1992ema} and uncertainty-based multi-task weighting~\cite{kendall2018multi}. Checkpoint selection, early-stopping criteria and any differences between the headline experiments and the data-efficiency study are defined in the released scripts.

\subsection{\label{app:data_splits}Data splits and data-efficiency protocol}

The main \fasercal{} experiments use one canonical 85/5/10 train/validation/test split. The data-efficiency study reuses the same validation and test partitions and subsamples only the training partition at budgets of 100, 300, 1,000, 3,000, 10,000, 30,000 and 100,000 events, with three seeds per budget. Because the pooled \fasercal{} sample contains the dedicated $\nu_\tau$-enriched component described in Methods, flavour-composition-sensitive test-set metrics are reweighted so the $\nu_\tau$ abundance matches the nominal unbiased sample.

The scintillator transfer study uses the public training/testing split and reserves 5\% of the public training split for validation with seed 42. The PILArNet study uses a fixed 80k/2k/18k train/validation/test event split on the public $768^3$-pixel release. Detector-spanning geometric augmentation is intentionally limited in the \fasercal{} case, because the detector subsystems are sequential and not spatially aligned.

\section{\label{app:losses}Supplementary target and loss definitions}

\subsection{\label{app:pretrain_targets}Pre-training targets}

Masked reconstruction predicts voxel occupancy and charge for masked \threedcal{} and \ahcal{} patches, together with masked \ecal{} and muon-spectrometer summaries when those inputs are dropped. The relational pass predicts ghost labels (binary), hierarchy labels (three classes: background, primary, secondary) and particle-category labels (three classes: electromagnetic, muonic, hadronic) on kept \threedcal{} voxels. For the semantic targets, labels are not one-hot: each reconstructed voxel inherits all matched truth contributions, weighted by their fractional contribution to that voxel and normalised across classes, so several classes can coexist in a single voxel. Ghost remains a hard binary target. The reconstruction losses use a hybrid formulation that combines standard voxel-level terms with soft-chamfer and distance-weighted regression components, providing smoother gradients for near-miss predictions around sparse shower boundaries.

\subsection{\label{app:finetune_targets}Fine-tuning targets}

Fine-tuning jointly predicts a six-way flavour label, a four-way charm label, visible momentum, jet momentum, and the primary vertex. The $\nu_\tau$ classes are defined from the primary tau-decay products: $\nu_\tau$ CC$\to e$ and $\nu_\tau$ CC$\to \mu$ require an electron or muon, respectively, while all remaining $\nu_\tau$ CC events are assigned to $\nu_\tau$ CC$\to \mathrm{had}$. Charm labels are defined analogously from the primary charm-decay products: charm$\to \mu$ takes precedence, followed by charm$\to e$, with the remainder assigned to charm$\to \mathrm{had}$. The regression heads predict the visible and hadronic-jet momentum vectors, parametrised in cylindrical coordinates $(p_T, \phi, p_z)$ with a log-space transform for magnitudes. In the analysis code, the reported scalar observables $E_{\mathrm{vis}}$ and $p_T^{\mathrm{miss}}$ are derived from the visible momentum, while the primary-lepton momentum is constructed as $\vec{p}_\ell=\vec{p}_{\mathrm{vis}}-\vec{p}_{\mathrm{jet}}$. The reported quantities $|p_\ell|$ and $|p_{\mathrm{jet}}|$ denote the magnitudes of the primary-lepton and hadronic-jet momenta, and the vertex metric is $d_{\mathrm{PV}}=\|\vec{x}_{\mathrm{PV}}^{\mathrm{true}}-\vec{x}_{\mathrm{PV}}^{\mathrm{reco}}\|$. All task losses are combined through learned homoscedastic uncertainty weights~\cite{kendall2018multi}.

\end{document}